\documentclass{emulateapj}

\shorttitle{The rotating accretion disk surrounding NGC\,7538\,S}
\shortauthors{Sandell et al.}
\slugcomment{Accepted by ApJ}

\begin{document}

\def\arcmin{{$^{\prime}$}}
\def\arcsec{{$^{\prime\prime}$}}
\def\ptsec{$''\mskip-7.6mu.\,$}
\def\psec{$^s\mskip-7.6mu.\,$}
\def\Msun{\,{\rm M$_{\odot}$}}
\def\Lsun{\,{\rm L$_{\odot}$}}
\def\ltsim{$\stackrel{<}{\sim}$}
\def\gtsim{$\stackrel{>}{\sim}$}
\def\jtra#1#2{${\rm J}\!\!=\!\!{#1}\!\!\to\!\!{#2}$}

\def\degr{$^{\circ}$}

\title{A detailed study of the accretion disk surrounding the high-mass protostar NGC\,7538\,S}

\author{G\"oran Sandell\altaffilmark{1}, Melvyn Wright\altaffilmark{2}}

\altaffiltext{1}{SOFIA-USRA, NASA Ames Research Center, MS N211-3,
Moffett Field, CA 94035, USA}
\email{Goran.H.Sandell@nasa.gov}
\altaffiltext{2}{Radio Astronomy Laboratory, University of California, Berkeley,
601 Campbell Hall, Berkeley, CA 94720, USA}

\begin{abstract}
We present deep high angular resolution observations  of the high-mass
protostar NGC\,7538\,S, which is in the center of a cold dense cloud
core with a radius of 0.5 pc and a mass of $\sim$ 2,000 \Msun. These
observations show  that NGC\,7538\,S is embedded in a compact elliptical
core with a mass of 85 - 115 \Msun. The star is surrounded by a rotating
accretion disk, which powers a very
young, hot molecular outflow approximately perpendicular to the rotating
accretion disk. The  accretion rate is very high, $\sim$ 1.4 --  2.8
10$^{-3}$ M$_{\odot}$~yr$^{-1}$. Evidence for rotation of the disk
surrounding the star is seen in all  largely optically thin molecular tracers,
H$^{13}$CN J = $1 \to 0$, HN$^{13}$C J = $1 \to 0$, H$^{13}$CO$^+$ J =
$1 \to 0$, and DCN J = $3 \to 2$.  Many molecules appear to be affected
by the hot molecular outflow, including DCN and H$^{13}$CO$^+$. The
emission from CH$_3$CN,  which has often been used to trace disk
rotation in young high-mass stars, is dominated by the outflow,
especially at higher K-levels. Our new high-angular resolution
observations show that the rotationally supported part of the disk is
smaller than we previously estimated. The enclosed mass of the inner,
rotationally supported part of the disk (D $\sim$ 5\arcsec{}, i.e 14,000
AU) is $\sim$  14 - 24\Msun. 

\end{abstract}


\section{Introduction}
\label{Intro}

How high-mass stars are formed is still under debate. High-mass stars
are believed to form the same way as low mass stars, i.e. with a
rotating accretion disk and driving an outflow, but in denser
environments and with much higher accretion rates \citep{Wolfire87,
Stahler00,Norberg00,McKee02,Keto03,Keto07}. However, others argue that
high-mass stars are most likely formed by competitive accretion in a
clustered environment \citep{Bonnell06}, or by adiabatic accretion of
gas from the surrounding cluster to densities such that stellar
collisions are likely to occur \citep{Clarke08}.

Accretion disks are ubiquitous in young low- and intermediate-mass
stars, i.e. T Tauri and Herbig Ae stars
\citep{Simon00,Mannings97,Najita03}, which have sizes of one to a few
hundred AU and appear to be in Keplerian rotation. Disks are also common
in Class I objects, i.e. younger, more heavily embedded low- and
intermediate mass pre-main-sequence stars \citep{Brown99}, but there are
very few examples of confirmed disks for the very earliest stages of  a
low-mass protostar, the Class 0 phase \citep{Jorgensen04,Chandler05}. At
the Class 0 phase, the temperature of the protostellar disk is of the
same order of that of the collapsing cloud core, and it is therefore an
observational challenge  to get enough contrast between the disk and the
cloud core.

The situation is even worse for high-mass stars, because they form in
much denser environments and always in clusters or small groups. The
survival time of disks in high-mass stars is furthermore expected to be
much shorter, because once the star is formed it will quickly photo
evaporate the disk as soon as the accretion from the surrounding cloud
starts to diminish, therefore allowing the uv-photons to reach the disk
surface. Yet there are several detections of disks around early B-stars;
Lkh$\alpha$\,101 (Spectral class B0) \citep{Tuthill02}, MWC297 (B1.5)
\citep{Manoj07}, MWC\,349\,A (B [e]) \citep{Weintroub08}, clearly
confirming that disks do exist around relatively massive stars.

Most young high-mass stars appear to drive outflows
\citep{Shepherd96,Zhang01,Zhang05,Beuther02b}, but whether they also
have accretion disks is less clear.  There have been many reports of
disks around high-mass stars, but only a few which appear well
supported. Two of the most clearcut cases are IRAS
$20126+4104$\citep{Cesaroni97,Cesaroni99, Cesaroni05,Zhang98}, and
AFGL\,490 \citep{Schreyer06}. IRAS $20126+4104$ has a
luminosity of $\sim$ 1.3 10$^4$ \Lsun\ suggesting an early B star. The
star drives a parsec scale CO  outflow \citep{Wilking90,Shepherd00}, and
is surrounded by a rotating accretion disk with a diameter of $\sim$
15,000 A.U. \citep{Cesaroni05}. The observed rotation curve for an
edge-on Keplerian disk corresponds to $\sim$ 7 \Msun. The luminosity of
AFGL\,490, L $\sim$  2 10$^3$ \Lsun\ also suggests an early B-star. It
drives a  more compact outflow, 0.3~pc (corrected for inclination)
\citep{Lada81,Snell84} and is surrounded by a nearly face-on disk with a
radius  $\sim$ 1600 AU, which has  a disk mass of 1 \Msun\  and appears
to be in Keplerian rotation \citep{Schreyer06}. Another, even younger
high-mass protostar, NGC\,7538\,S, with a luminosity of $\sim$ 1.5~10$^4$
\Lsun,  which  is surrounded by a large, nearly edge-on disk with a
radius of 15,000 AU and a disk mass of $\sim$ 100 \Msun\ was reported by
\citet{Sandell03}. The disk drives a compact molecular outflow with a
size of 0.1 - 0.4 pc and shows a Keplerian-like rotation.

This disk is the target for our studies. In this paper we
present follow-up BIMA spectral line observations of several molecules
in the 3 and 1 mm bands, which were chosen to measure the rotation of
the disk surrounding NGC\,7538\,S. In a  companion paper
(Corder et al., in preparation; hereafter Paper II) we discuss high-angular resolution
continuum observations of NGC\,7538\,S and the Ultra-Compact \ion{H}{2}
regions NGC\,7538 IRS\,1 - 3 with BIMA, CARMA, and the VLA, which
provide us with better mass estimates of the disk/envelope surrounding
the star.

\section{Observations and Data Reduction}

\subsection{BIMA array observations}
\label{BIMA}

Between 2002 and 2004 we obtained more observations on NGC\,7538\,S
with the BIMA array at  86 and 89 GHz. These observations are summarized
in Table \ref{tbl-1}. At 3.4~mm (87~GHz) we combined additional
data obtained in the B, C and D array configurations with the data
reported by \citep{Sandell03} to produce deep high fidelity maps in
H$^{13}$CN \jtra10, HCO$^+$ \jtra10, SO $2_2 \to 1_1$ and NH$_2$D
1$_{11}  \to  1_{01}$. We also improved the H$^{13}$CO$^+$ \jtra10,
HN$^{13}$C \jtra10, and SiO \jtra21, by adding a B-array track to the
existing C-array data. All these observations use the same correlator
setup and were reduced as described in  \citet{Sandell03}.

Table \ref{tbl-2} summarizes the spectral line observations at 1.4~mm
(220 GHz), which we obtained with BIMA during the same time frame
as the 3.4~mm observations. These 1.4~mm observations were obtained in the B
and C configurations, and provide similar angular resolution as the 87
GHz data sets.  In this paper we discuss the results from DCN J = $3 \to 2$ 
and CH$_3$CN J = $12 \to 11$ (K = 0 to 5), which
were expected to  be optically thin and hence good tracers of the accretion disk. 
Our observations
also included SO  J = $5_5 \to 4_4$, H$_2$CO   3$_{03}  \to 2_{02}$ and 
3$_{22}  \to 2_{21}$, and  $^{13}$CO J = $2 \to 1$ (Table \ref{tbl-2}),
but the lines are optically thick and self-absorbed and do not provide
us any information on the deeply embedded accretion disk. We therefore
leave the discussion of these lines to a later paper.

The 1~mm data were reduced and imaged in a standard way using MIRIAD
software \citep{Sault95}. The quasar 0102$+$584 was used as phase
calibrator and Mars and/or 3C84 for passband and flux calibration.  The
uncertainty in the absolute amplitude scale is $\sim$ 20\% at 1.4~mm,
but the relative amplitude of the spectral lines observed simultaneously
is $\sim$5 \%.

The BIMA array, like every interferometer, will filter out extended,
relatively uniform emission, resulting in an underestimate the
total emission (missing flux). Furthermore, extended emission will
cause negative features in the spectra, which may look like
self-absorption. Since we expect to see self-absorption if the
protostellar disk is warmer than the surrounding cloud, it becomes
difficult, if not impossible, to judge what features are real, and what
are artifacts, unless we fill in missing short uv-spacings. We
therefore carried out observations with FCRAO at 3.4~mm band and with
JCMT at 1.4 mm to provide missing short spacings for several key molecules.


\begin{deluxetable}{lll}
\tablecolumns{3}
\tablenum{1}
\tablewidth{0pt} 
\tablecaption{BIMA 3 mm observing log.} 
\label{tbl-1}
\tablehead{
\colhead{Molecule} & \colhead{transition} & \colhead{rest frequency} \\ 
\colhead{}  &\colhead{}  & \colhead{[GHz]}  \\
\cline{1-3} 
}
\startdata
\cline{1-3}
\sidehead{Frequency setting: H$^{13}$CN \jtra10; Array configurations BCD, } 
\sidehead{\phantom{Frequency setting}HPBW = 6\ptsec0 $\times$ 5\ptsec7 pa = -25.8\degr{}}
\sidehead{Observing dates:  01/27/02, 01/31/02, 05/03/02, 08/09/03,}
\sidehead{\phantom{Observing dates: }11/04/03, 12/11/0312/12/03, 12/31/03} 
\cline{1-3}
NH$_2$D\phantom{XXXX} & 1$_{11}  \to  1_{01}$  F$^{\prime}$ - F$^{\prime\prime}$ = 0 - 1& 85.924747  \\
NH$_2$D & 1$_{11}  \to  1_{01}$  F$^{\prime}$ - F$^{\prime\prime}$  = 2 - 1 & 85.925684 \\
NH$_2$D &  1$_{11}  \to  1_{01}$  F$^{\prime}$ - F$^{\prime\prime}$ = 2 - 2/1 - 1 & 85.926263\\
NH$_2$D  &  1$_{11}  \to  1_{01}$  F$^{\prime}$ - F$^{\prime\prime}$ = 1 - 2 & 85.926858 \\
NH$_2$D &  1$_{11}  \to  1_{01}$  F$^{\prime}$ - F$^{\prime\prime}$ = 1 - 0 & 85.927721  \\

SO  & \jtra{2,2}{1,1} & 86.09355 \\
H$^{13}$CN & \jtra10 F = 1 - 1 & 86.338767 \\
H$^{13}$CN & \jtra10 F = 2 - 1 & 86.340184 \\
H$^{13}$CN  &\jtra10 F = 0 - 1 & 86.342274 \\
HCO$^+$  &\jtra10 & 89.188518 \\
\cline{1-3}
\sidehead{Frequency setting: H$^{13}$CO$^+$ \jtra 10; Array configurations BC, }
\sidehead{\phantom{Frequency setting}HPBW = 7\ptsec9 $\times$ 6\ptsec9 pa = -23.8\degr{}}
\sidehead{Observing dates: 12/03/01, 01/10/04 } 
\cline{1-3} 
H$^{13}$CO$^+$& \jtra10 & 86.754294 \\
SiO  & \jtra21 v=0  &  86.846998  \\
HN$^{13}$C  & \jtra10 F = 0-1 & 87.090735 \\
HN$^{13}$C  & \jtra10 F = 2-1 & 87.090859 \\
HN$^{13}$C  &  \jtra10 F = 1-1 & 87.090942 \\
\enddata
\end{deluxetable}

\begin{deluxetable}{llll}
\tablecolumns{3}
\tablenum{2}
\tablewidth{0pt} 
\tablecaption{BIMA 1 mm observing log.} 
\label{tbl-2}
\tablehead{
\colhead{Molecule} & \colhead{transition} & \colhead{rest frequency} \\ 
\colhead{}  &\colhead{}  & \colhead{[GHz]}  \\
}
\startdata
\sidehead{Frequency setting: H$_2$CO 3$_{03}  \to 2_{02}$; array configurations: BC}
\sidehead{\phantom{Frequency setting:  }HPBW = 6\ptsec6 $\times$ 5\ptsec7 pa = 78.4\degr{}}
\sidehead{Observing dates:  10/9/02, 10/16/02} 
\cline{1-3}
SO  & \jtra{5,5}{4,4} &\phantom{1111111111} 215.22065 \\
H$_2$CO & 3$_{03}  \to 2_{02}$  &\phantom{1111111111}  218.22219 \\
H$_2$CO & 3$_{22}  \to 2_{21}$  &\phantom{1111111111}  218.47564  \\
\cline{1-3}
\sidehead{Frequency setting: DCN \jtra32;  array configurations: BC}
\sidehead{\phantom{Frequency setting:  }HPBW = 2\ptsec9 $\times$ 2\ptsec4 pa = 22.2\degr{}}
\sidehead{Observing dates: 10/13/02, 04/19/03, 05/01/03, 10/10/03,}
\sidehead{\phantom{Observing dates: }10/25/03, 01/03/04, 01/05/04}
\cline{1-3}
DCN    &   \jtra32 &\phantom{1111111111}   217.23853   \\
$^{13}$CO    &   \jtra 21 &\phantom{1111111111}   220.39869  \\
CH$_3$CN & 12$_0 \to 11_0$ &\phantom{1111111111}  220.74727  \\
CH$_3$CN & 12$_1 \to 11_1$ &\phantom{1111111111}  220.74302  \\
CH$_3$CN & 12$_2 \to 11_2$ &\phantom{1111111111}  220.73027  \\
CH$_3$CN & 12$_3 \to 11_3$ &\phantom{1111111111}  220.70903  \\
CH$_3$CN & 12$_4 \to 11_4$ &\phantom{1111111111}  220.6793  \\
CH$_3$CN & 12$_5 \to 11_5$ &\phantom{1111111111}  220.64109 \\
CH$_3$CN & 12$_6 \to 11_6$ &\phantom{1111111111}  220.59443  \\
CH$_3$$^ {13}$CN  & 12$_2 \to 11_2$ &\phantom{1111111111}  220.62108  \\
CH$_3$$^ {13}$CN  & 12$_3 \to 11_3$ &\phantom{1111111111}  220.59994 \\
\enddata
\end{deluxetable}

\begin{figure}
\includegraphics[ scale=0.49,angle=-90]{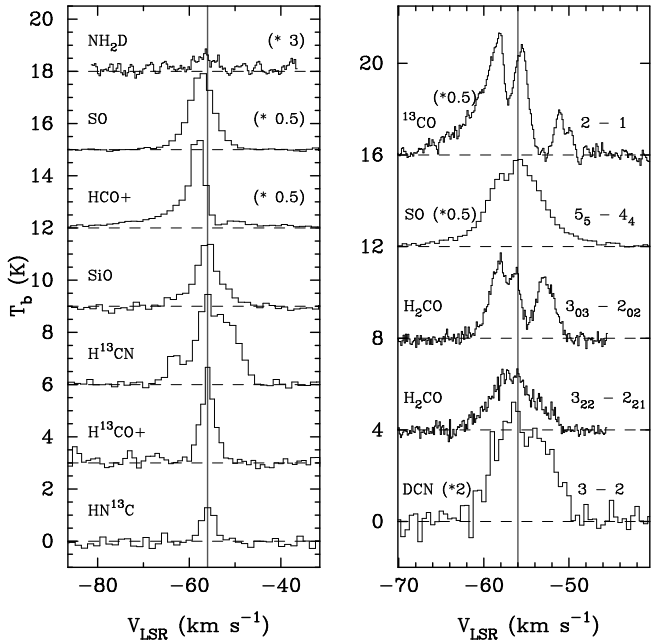}

\figcaption[]{
\label{fig-spectra}
Spectra  toward NGC\,7538\,S  at 3~mm (left panel) and 1~mm (right
panel). Spectra for HCO$^+$ and H$^{13}$CN in the 3~mm window and  for
$^{13}$CO and H$_2$CO in the 1~mm window are BIMA data merged with
single dish data to correct for missing zero-spacings. The red-shifted
self-absorption we see in these molecular transitions is therefore real
and not due to missing flux filtered out by the interferometer. The grey
vertical line marks the systemic velocity. In the 3~mm window HCO$^+$,
SiO and SO show strong line wings; both SO and HCO$^+$ are self-absorbed
at red-shifted velocities. At 1~mm all the spectra shown here exhibit
line wings. $^{13}$CO \jtra21 and H$_2$CO   3$_{03} \to 2_{02}$ show
deep self-absorption at red-shifted velocities. Even DCN \jtra32 is
affected by some red-shifted self-absorption. The scaling factor is
shown in parenthesis.
}
\end{figure}

\subsection{FCRAO observations}

The observations were carried out on November 9, 2004 by Dr. Mark Heyer
with the 14-m Five College Radio Astronomy Observatory (FCRAO) telescope
using the SEQUOIA 16 beam array receiver. The weather conditions were
excellent, resulting in a system temperature, T$_{sys}$ $\sim$ 120~K. We
used the dual channel correlator (DCC) with a 50 MHz bandwidth and a
spectral resolution of 48.8  kHz ( 0.16 km~s$^{-1}$). The DCC was
configured to simultaneously cover  H$^{13}$CN \jtra10 and HCO$^+$
\jtra10.  At 86 and 89 GHz the half-power beam width (HPBW) is
58\arcsec\ and 57\arcsec, respectively, while the main-beam efficiency,
$\eta_{\rm mb} \sim$ 0.50. The observations were performed using
on-the-fly (OTF) mapping, covering a 9\arcmin\ $\times$ 9\arcmin\ area
centered at \mbox{$\alpha_{2000.0}$ = 23$^h$ 13$^m$ 45\psec4},
\mbox{$\delta_{2000.0}$ = $+$61\degr{} 28\arcmin{} 10\ptsec5}, and
covering the whole molecular cloud associated with NGC\,7538. The OTF
maps were re-gridded to a cell size 20\arcsec.  Both data sets were
merged with our BIMA data using the MIRIAD task IMMERGE and assuming the
HPBWs and efficiencies for FCRAO quoted above. The single dish and
interferometer  data were compared in the Fourier domain in a range of
overlapping spatial frequencies corresponding to 6-10m  to align the
flux calibration scales.

\subsection{JCMT observations}

Complementary single dish observations  (projects m04ai09 and m05ai07)
were carried out in service mode on the James Clerk Maxwell Telescope
(JCMT)\footnote{The JCMT is operated by the Joint Astronomy Centre,
on behalf of the UK Particle Physics and Astronomy Research Council,
the Netherlands Organization for Scientific Research, and the Canadian
National Research Council.} on Mauna Kea, Hawaii, between April 2004
and May 2005. With JCMT we obtained fully sampled maps of DCN \jtra32,
H$_2$CO   3$_{03}  \to 2_{02}$ and 3$_{22}  \to 2_{21}$, $^{13}$CO
\jtra21, and C$^{18}$O \jtra21, as well as long integration (10 - 30 min)
single spectra at the center position (NGC\,7538\,S). We also took deep
(long integration) spectra of $^{12}$CO and  C$^{17}$O \jtra21. All these
observations used receiver RxA3, a single beam SiS mixer covering the
1.3~mm spectral window. At 217 GHz the measured HPBW is 22\ptsec3 and
the main beam efficiency, $\eta_{\rm mb}$ = 0.70. All maps were made
in OTF-mode with heavy oversampling; 5\arcsec\ sampling interval  with
an integration time of 5 second per resolution element. The maps have a
size 55\arcsec\ $\times$ 45\arcsec. We used the Digital Autocorrelator
Spectrometer (DAS) with 125 MHz bandwidth ($^{13}$CO and C$^{18}$O) or
250 MHz bandwidth (DCN, H$_2$CO) providing us with spectral resolutions
78.1 kHz and 156.2 kHz, respectively. Most of the observing was in
wet weather conditions, resulting in system temperatures, T$_{sys}$ of 400
- 450~K. The receiver calibration was regularly checked on planets (Uranus
or Mars) and we always took reference spectra on NGC\,7538\,IRS\,1.

Our DCN setup also included the SiO \jtra43 transition at 217.10494
GHz. In total we acquired 10 maps, because the line is rather faint. Some
scans had poor baseline stability and had to be thrown out, and
for most of the observations we had to use high order polynomials
for baseline subtraction. During the data reduction we discovered an
additional line in our spectra, which we identified as the SO$_2$
13$_{2,12} \to 13_{1,13}$ at 225.153689 GHz, i.e. in the upper
sideband. SiO and SO$_2$ show broad line wings, indicating that the
emission is strongly affected by the outflow. The DCN-map shows that
the emission is rather compact  with a size $\sim$ 30\arcsec. When we
add it to the BIMA data, we do not see any changes in the morphology
of the DCN emission. However, since the combined map had poorer signal
to noise due to insufficient integration time, we make use of only the
high-angular resolution BIMA map in this paper.

Both H$_2$CO transitions were observed simultaneously. We obtained
three maps with good quality. The coadded map matched well the noise
level in our BIMA data. At $^{13}$CO \jtra21 we obtained a
total of 7 maps, largely because we needed good SNR on the faint wings
from the outflow. The coadded map was of very good quality and was
successfully added to our BIMA data on  $^{13}$CO. For C$^{18}$O
\jtra21 we obtained  4 maps, also of good quality. Since we never mapped
NGC\,7538\,S  with BIMA in C$^{18}$O, we have used this map stand-alone
to look at the large scale morphology of the cloud core.  

\begin{deluxetable*}{lccrcc}
\tablecolumns{6}
\tablenum{3}
\tablewidth{0pt} 
\tablecaption{Gaussian fits with CLASS of optically thin or moderately optically thin lines towards the center of the accretion disk, \mbox{$\alpha_{2000.0}$ = 23$^h$ 13$^m$ 44\psec98},
\mbox{$\delta_{2000.0}$ = $+$61\degr{} 26\arcmin{} 49\ptsec7}.}

\label{tbl-3}
\tablehead{
\colhead{Molecule} & \colhead{Transition}& \colhead{HPBW} &\colhead{$\int T_{\rm MB} dV$} & \colhead{$\Delta V$} & \colhead{$V_{\rm LSR}$} \\
\colhead{ }         &\colhead{ }        &\colhead{[ \arcsec{} $\times$ \arcsec{}] }   & \colhead{[K km s$^{-1}$]}& \colhead{[km s$^{-1}$]}& \colhead{[km s$^{-1}$]} \\
}
\startdata
 C$^{18}$O &$2 \to 1$&  21.6 $\times$ 21.6   &  33.44 $\pm$ 0.21  &  4.68 $\pm$ 0.03  &  $-$56.42 $\pm$ 0.01  \\
 $\phantom{c18o} $\tablenotemark{a}              & &  &  $-$1.22 $\pm$ 0.11 & 2.10 $\pm$ 0.20 &  $-$55.73 $\pm$ 0.11 \\         
 C$^{17}$O & $2 \to 1$ &    21.2 $\times$ 21.2 &       9.53  $\pm$ 0.03& 5.04 $\pm$  0.18 & $-$56.42 $\pm$ 0.02  \\
 H$^{13}$CO$^+$  & $1 \to 0$&  7.9 $\times$ 6.9 &       14.73  $\pm$ 1.00 &   4.20 $\pm$ 0.36   &   $-$55.96 $\pm$ 0.14    \\
 HN$^{13}$C\tablenotemark{b}  & $1 \to 0$ & 7.8 $\times$ 7.1 & 5.40  $\pm$ 0.17 &  3.60  $\pm$ 0.10 &  $-$55.91 $\pm$ 0.36  \\
SiO\tablenotemark{c}                    & $2 \to 1$&   8.2 $\times$ 7.1  &  7.85  $\pm$ 0.90 & 3.73 $\pm$ 0.33 & $-$56.23 $\pm$ 0.13\\
H$^{13}$CN\tablenotemark{d,e}  & $1 \to 0$& 5.8 $\times$ 5.7 & 37.00  $\pm$ 0.70  &  4.70  $\pm$ 0.10&  $-$56.20 $\pm$ 0.13  \\
H$^{13}$CN\tablenotemark{f}  & $1 \to 0$& 4.0 $\times$ 3.7 & 24.86  $\pm$ 0.22 &  5.85  $\pm$ 0.05 &  $-$55.97 $\pm$ 0.12 \\
NH$_2$D\tablenotemark{f}  & 1$_{11} \to 1_{01}$& 12.1 $\times$ 11.9 & 5.30  $\pm$ 0.06 &  1.35  $\pm$ 0.10 &  $-$56.28 $\pm$ 0.06  \\ 
DCN & $3 \to 2$ &  2.9 $\times$ 2.4 & 21.69 $\pm$ 0.84 &  6.89 $\pm$ 0.18 & $-$55.56 $\pm$ 0.18 \\
 $\phantom{DCN}$\tablenotemark{a}     & &   & $-$2.88 $\pm$ 0.70 &  2.65 $\pm$ 0.01 & $-$54.85 $\pm$ 0.21 \\
\enddata
\tablenotetext{a}{Fit to self-absorption component }
\tablenotetext{b}{$\Delta V$ and $V_{\rm LSR}$ from method hfs in CLASS, see Section~\ref{Results}.}
\tablenotetext{c}{This is  the cloud component only. SiO is dominated by high velocity emission, which has a  line integral of  16.6  K km s$^{-1}$ and a line width of 13.6 km s$^{-1}$.}
\tablenotetext{d}{Single dish corrected spectrum}
\tablenotetext{f}{Fit with  hyperfine components locked in velocity separation to the main hyperfine line. }
\tablenotetext{f}{High spatial resolution BIMA spectrum; $V_{\rm LSR}$ from method hfs.}
\end{deluxetable*}

\section{Results and Analysis}
\label{Results}

In Figure \ref{fig-spectra}{} we show spectra towards NGC\,7538\,S of
all the molecular transitions, which have been observed with BIMA in the
3~mm and 1~mm band, except  the CH$_3$CN J = 11 $\to$ 10 lines, which we
will cover in Section \ref{mcn}. Although some of the changes in the
line profiles are due to  different  spatial resolution, this effect is
rather minor, especially since most of the spectral lines have been
observed with roughly the same angular resolution (Table \ref{tbl-1} \&
\ref{tbl-2}). Most of the differences in line shape results from whether
the lines are optically thick or thin, and whether they originate from
the disk, the outflow, or the cloud envelope. Almost all the lines
observed in the 1~mm band are optically thick, have strong line wings,
and show deep self-absorption from the surrounding cloud envelope. The
same is also true for HCO$^+$ \jtra10, SiO \jtra21, and SO
\jtra{2,2}{1,1} in the 3~mm band. In this paper we are focusing on the
accretion disk. We therefore limit our analysis to high density tracers,
which have dipole moments of 3 Debye or larger, and  which can be
assumed to be optically thin or at least have only moderate optical
depths.  Of the molecules we have observed, H$^{13}$CN, HN$^{13}$C, DCN,
H$^{13}$CO$^+$, CH$_3$$^{13}$CN, and CH$_3$CN, all fall into this
category.  These molecules are expected to be good tracers of the gas in
the dense disk surrounding NGC\,7538\,S. Most of them are isotopomers,
which have molecular abundance ratios of $\lesssim$ 10$^{-10}$ compared
to H$_2$. One would therefore expect them to be optically thin, but as
we will see later, this is not the case for all of the lines. Several of
them are also affected by emission from the molecular outflow. We have
therefore carefully inspected each image to see whether the line
emission originates from the disk, the outflow,  or whether both the
disk and the outflow contribute to the emission. The gas in the
surrounding cloud core is also very dense, and colder than the gas in
the disk. It can therefore absorb some of the emission from the disk.
However, since the surrounding envelope is much more extended than the
disk and relatively uniform, it is largely filtered out in these
interferometer observations, but may nevertheless add some contribution
to the line profiles observed towards the accretion disk.

In order to determine the systemic velocity of the disk, we extracted
spectra from our BIMA images for  all optically thin or moderately
optically thin molecules (see Table~\ref{tbl-3}) at the center of the
accretion disk. These spectra were imported into the IRAM single dish
spectral line reduction package CLASS, which is part of the GILDAS
software suite\footnote{http://www.iram.fr/IRAMFR/GILDAS}. CLASS has
very good line fitting routines and can also handle molecules with
hyperfine splitting. The ``method hfs'' allows fitting for optical depth
of lines with hyperfine structure, and one can derive more accurate
radial velocities by locking the velocity separation for satellite lines
to the main emission component as long as all the spectral transitions
are observed within a single band. We also fitted deep, single dish JCMT
spectra of the CO isotopomers C$^{18}$O, and C$^{17}$O J = $2 \to 1$ to
determine the systemic velocity of the cloud.The results of these
Gaussian fits are given in Table \ref{tbl-3}.  C$^{18}$O \jtra21 and DCN
\jtra32 show a narrow self-absorption feature at the cloud velocity or
at slightly red-shifted velocities, suggesting that both have an optical
depth of at least one in the surrounding cloud core. H$^{13}$CN \jtra10
shows evidence for self-absorption in the main hyperfine component at
high angular resolutions, see Section \ref{h13cn}. Even C$^{17}$O J = $2
\to 1$ shows a weak hint of self-absorption and a fit to the hyperfine
splitting of the molecule (method hfs) gives an optical depth $\tau
\sim$ 0.25. The results from Table  \ref{tbl-3} suggest a systemic
velocity of $-$56.0 $\pm$ 0.2 km~s$^{-1}$ . This agrees very well with
what we derive from the rotation curve of  H$^{13}$CN \jtra10, while the
DCN rotation curve would suggest a slightly more red-shifted systemic
velocity,  $-$55.7~km~s$^{-1}$.

\subsection{The NGC\,7538\,S star forming cloud core}
\label{core}

\begin{figure*}
\includegraphics[ scale=0.77,angle=-90]{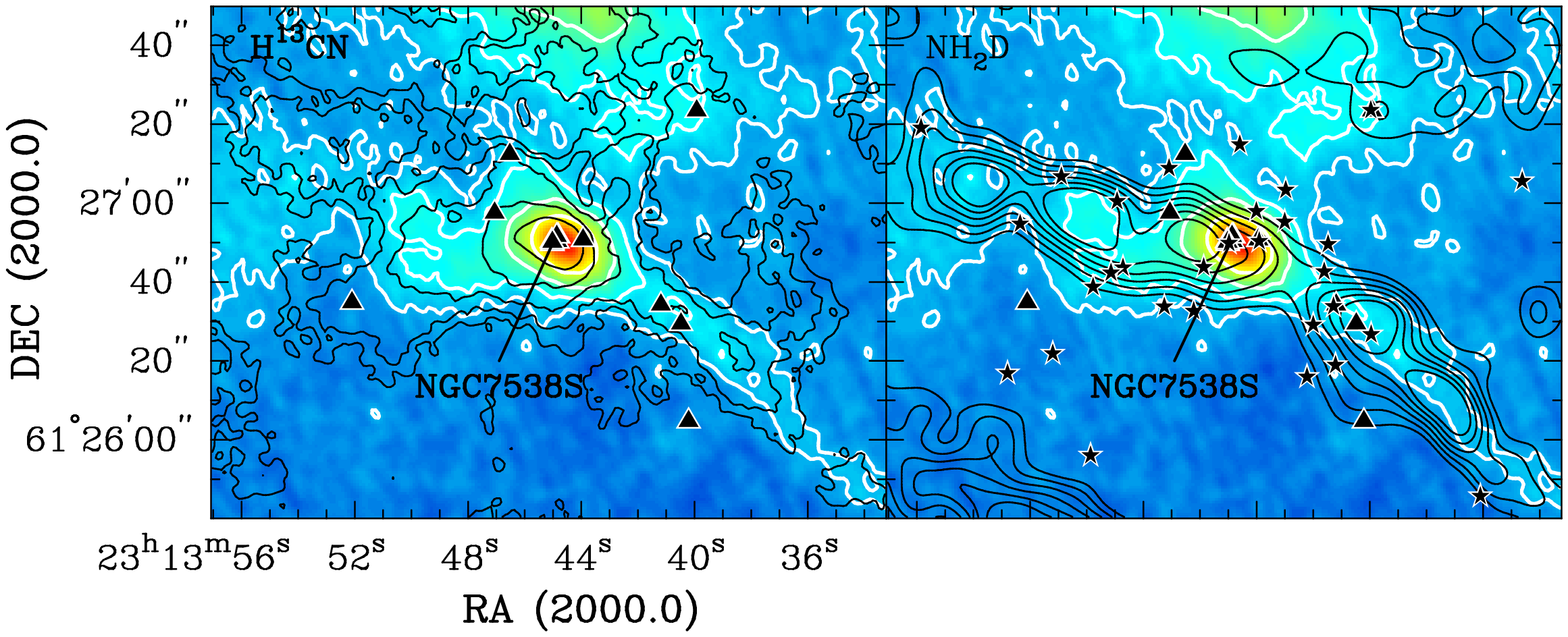}

\figcaption[]{
\label{fig-h13cn_nh2d_450}
Contour plots of total integrated H$^{13}$CN \jtra10 emission (left
panel) and of the NH$_2$D 1$_{11}  \to  1_{01}$ emission (right panel)
overlaid on a color image of  the 450 $\mu$m continuum emission. The
450 $\mu$m map is taken from \citet{Sandell04}. It has been cleaned and
restored with a 10\arcsec\ circular beam to better bring out the faint
low level dust emission. The peak flux of the 450 $\mu$m emission is
77.3 Jy~beam$^{-1}$. In order to better see the low level emission we
have enhanced the color image with four thick white contours plotted
logarithmically from 4 Jy~beam$^{-1}$ to half intensity. The H$^{13}$CN
emission is plotted with six logarithmic black contours, starting at 28
mJy~beam$^{-1}$  (3-$\sigma$) to peak intensity, 590 mJy~beam$^{-1}$. For
NH$_2$D we use an image with 12\arcsec\  resolution. The contours for
this image are plotted with  six black logarithmic contours going from  9
mJy~beam$^{-1}$ (3-$\sigma$) to peak intensity,  120 mJy~beam$^{-1}$. We
have overlaid the images with H$_2O$ masers (black triangles) and IRAC
8 $\mu$m sources (black star symbols) (Paper II), which pinpoint the
location of young stars in this region. We have also labeled the position
of NGC\,7538\,S.}
\end{figure*}

The NGC\,7538\,S star forming core is a cold, dense cloud core $\sim$
80\arcsec\ south of NGC\,7538 IRS\,1 \citep{Werner79,Zheng01,Sandell04}.
Here we have assumed a distance of 2.8 kpc for easy comparison with
earlier work, although recent trigonometric parallax observations
\citep{Moscadelli09}, result in a slightly smaller distance, 2.65 $\pm$
0.12 kpc. The star forming core has a luminosity of $\sim$ 10$^4$ \Lsun\
and harbors at least one massive protostar, NGC\,7538\,S
\citep{Sandell03}.  \citet{Zheng01}, who mapped the NGC\,7538 molecular
cloud in NH$_3$ with the VLA at high angular resolution, found the 
strongest ammonia emission towards NGC\,7538\,S, in good agreement with
the sub-mm continuum imaging by \citet{Sandell04}.  In the vicinity of
NGC\,7538\,S \citeauthor{Zheng01} found the NH$_3$ emission to be
optically thick with a temperature of $\sim$ 25 K.

Our H$^{13}$CN \jtra10 map  with FCRAO and BIMA is in good agreement
with the dust continuum and the NH$_3$ results. Overall we find
H$^{13}$CN to show the same cloud morphology as the dust continuum. This
is shown in Figure \ref{fig-h13cn_nh2d_450}, where we overlaid the
integrated H$^{13}$CN emission from our combined FCRAO and BIMA map on
the 450 $\mu$m SCUBA continuum  map  from \citet{Sandell04} for the
region around NGC\,7538\,S. The dust continuum and the H$^{13}$CN
emission both peak on NGC\,7538\,S and the narrow south-western filament
shows up clearly in both images. The dust emission appears somewhat more
extended than H$^{13}$CN to the east of NGC\,7538\,S, and there are also
some differences to the north, i.e. the southernmost part of the
NGC\,7538 IRS\,1 cloud core. Most of these differences are most likely
due to difference in temperature of the emitting gas and dust. Since
\citet{Zheng01} found very strong, cold NH$_3$ emission towards
NGC\,7538\,S, we decided to look for NH$_2$D emission as well, which we
could observe with the same frequency setup as H$^{13}$CN and HCO$^+$
(Section \ref{BIMA}, Table \ref{tbl-1}).  The NH$_2$D emission is quite
faint, especially towards NGC\,7538\,S, but  the narrow, lumpy,
southwestern filament shows up very well in the integrated NH$_2$D
emission (Figure  \ref{fig-h13cn_nh2d_450}b). However, the  NH$_2$D
emission is even stronger to the east of NGC\,7538\,S, where it shows a
``continuation'' of the southwestern filament, which is not evident in
dust continuum, nor is it seen in H$^{13}$CN or any other of the
molecules that we have imaged towards NGC\,7538\,S. Examination of
NH$_2$D spectra along the filaments show broadened lines at the
positions of the knots in the filaments and sudden, small velocity
shifts between the knots, suggesting that these ``clumps'' maybe
gravitationally unstable and that they may already have formed stars.
The only exception is the  NH$_2$D clump at the tip of the eastern
filament. This is where we find the  strongest  NH$_2$D emission, but
the lines are very narrow ($\Delta$v = 0.9 km~s$^{-1}$), suggesting that
this condensation is a very cold, prestellar core.

\begin{figure}
\includegraphics[ scale=0.57]{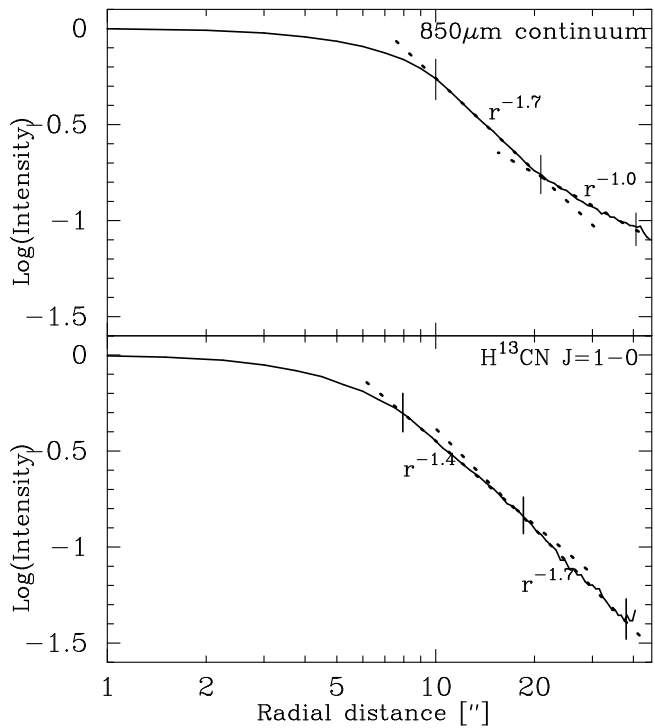}
\figcaption[]{
\label{fig-radial}
Radially averaged intensity profiles (normalized to one) of H$^{13}$CN J
= $1 \to 0$ and the 850 $\mu$m SCUBA image centered on NGC\,7538\,S. The
distribution of both H$^{13}$CN and dust emission is relatively flat at
small radii (r$<$ 8\arcsec{}). Between 8\arcsec\ and 20\arcsec\ the
falloff in density is proportional to r$^{-\alpha}$, where $\alpha$ is
in the range  1.4 - 1.7. At larger radii H$^{13}$CN falls off steeper,
while the dust emission flattens out. Both H$^{13}$CN and the dust
emission indicate a core size of $\sim$ 38\arcsec. The break points
between different regimes are shown by vertical lines in both plots.}
\end{figure}

We can use our high-quality spectral line and continuum data to estimate
the size and total mass of the NGC\,7538\,S star forming core.
\citet{Sandell04} quoted a radius  of 0.4 pc (30\arcsec{}) and a total
mass of  750 \Msun, assuming an average dust temperature of 30~K.  If we
use our combined FCRAO and BIMA H$^{13}$CN map and radially average the
data from the center of the proto-star, we find that the emission is
relatively flat at small radii,  has a steeper fall-off  to a radius of
$\sim$ 18 - 20\arcsec\ and  gets even steeper at larger radii until it
flattens off at $\sim$ 38\arcsec\ from the cloud center. The emission
stays roughly  constant at larger radii until it start to rise again
when we pick up emission from the IRS\,1 core.  If we radially average
the 450 $\mu$m map (Figure \ref{fig-h13cn_nh2d_450}), we find a core
size of 34\arcsec. The 850 $\mu$m image, however, has better sensitivity
than the 450 $\mu$m-map, and here the radially averaged dust emission
suggests a core radius of $\sim$ 40\arcsec, which agrees well with
H$^{13}$CN after correcting for the difference in HPBW. The dust
emission therefore has a similar distribution to H$^{13}$CN, although it
falls off less steeply at larger radii, see Figure~\ref{fig-radial}. 
Both dust and gas therefore predict a core radius of  $\sim$ 0.5~pc
(38\arcsec{}). If we assume an average gas temperature of 25~K,
consistent with the temperature derived from NH$_3$, and an abundance
ratio for H$^{13}$CN similar to the OMC-1 extended ridge
\citep{Blake87}, i.e.  [HCN]/[H$_2$] = 5 $\times$ 10$^{-9}$, and an
isotope ratio\footnote{The $^{12}$C/$^{13}$C  abundance is strongly
fractionated in cold cloud cores and $\sim$ 40 throughout OMC-1, see
e.g. \citet{Blake87}. It therefore appears more appropriate to use this
isotope ratio for the NGC\,7538 molecular cloud, rather than 85, which
was used by \citet{Sandell05} in their study of the IRS\,9 cloud core.
The  observed $^{12}$CO/$^{13}$CO ratios at high velocities suggest that
the isotope ratio could be slightly higher,  $\sim$ 60.}
[$^{12}$CO]/[$^{13}$CO] = 40, we obtain a total mass of $\sim$ 2,000
\Msun. 

As discussed above the core size is the same for  the dust emission as  found for
H$^{13}$CN, even though there are differences in the morphology on smaller scales.
 If we assume that the gas and dust is thermalized at 25~K , and
take a $\beta$-index of 1.5, the SCUBA maps give a total mass of $\sim$
2,000\Msun, which is in excellent agreement with H$^{13}$CN. The JCMT
C$^{18}$O \jtra21 map, however, gives a mass of only $\sim$ 1,000 \Msun,
if we assume that  C$^{18}$O is optically thin. We know that this
assumption is not true near the core center, and the total mass will
therefore be somewhat underestimated. However, the main reason for low
mass derived from CO is that a large fraction of the CO is  frozen onto
dust grains, resulting in a depletion of the CO gas phase abundance.
\citet{Mitchell90} found $\sim$ 10\% of the CO to be tied up into ice in
the NGC\,7538 IRS\,9  cloud core. Since the NGC\,7538\,S cloud core is
even colder than the IRS\,9 cloud, the fraction of solid to gas phase CO
is likely to be  higher. The low core mass deduced from C$^{18}$O is
therefore a result of CO depletion and optical depth. We therefore
conclude that the  total mass of the NGC\,7538\,S cloud core is $\sim$
2,000 \Msun, corresponding to an average gas density of $\sim$
6~10$^4$ cm$^{-3}$.

\subsection{The position of the central protostar, NGC\,7538\,S}

NGC\,7538\,S was detected in deep IRAC images with the {\em Spitzer}
Space Observatory at 4.5, 5.8, and 8 $\mu$m (Paper II). In paper II we
also present VLA observations, which show that the star drives a highly
collimated thermal jet, with the strongest emission knot coinciding with
the IRAC position to within 0\ptsec1. In this paper we have therefore
adopted the VLA position of NGC\,7538\,S as: \mbox{$\alpha_{2000.0}$ =
23$^h$ 13$^m$ 44\psec98}, \mbox{$\delta_{2000.0}$ = $+$61\degr{}
26\arcmin{} 49\ptsec7}  \footnote{Recent sub-arcsecond continuum imaging
at 110 and 224 GHz with CARMA (Paper II) resolve the elliptical core
into three compact sources, all of which are almost certainly
protostars.  The strongest one of the three sources agrees within
0\farcs15 with the adopted position for the high-mass protostar.}. This
position coincides within errors with the strongest OH maser spots
\citep{Argon00}, the position of the Class II CH$_3$OH maser
\citep{Pestalozzi06}, as well as with the dominant cluster of H$_2$O
masers \citep{Kameya90}. Furthermore, it places NGC\,7538\,S  on the
symmetry axis of the compact bipolar outflow imaged in HCO$^+$
\jtra{1}{0} and at the dynamical center of the rotating disk (Section
\ref{h13cn}).

\subsection{Outflow and Accretion}

We will discuss the molecular outflows in the NGC\,7538
molecular cloud in more detail in (Corder et al, in
preparation). In this section we re-examine the outflow, since it is
intimately connected to the thermal jet and both must be powered by
the accretion disk. The outflow parameters derived by \citet{Sandell03}
are incorrect, because they were derived from HCO$^+$ \jtra10 
observations  with the assumption that the outflow is cold and that the
HCO$^+$ abundance is the same as in quiescent molecular cloud cores.
However, \citet{Sandell05} analyzed the outflow from NGC\,7538~IRS\,9
and showed that the HCO$^+$ abundance is enhanced by more than a factor
of 30 compared to the ``standard'' OMC-1 ridge abundance. It is
therefore clear that the HCO$^+$ abundances are likely to be enhanced in the 
NGC\,7538\,S outflow as well. 

\begin{figure}
\includegraphics[ scale=0.39,angle=-90]{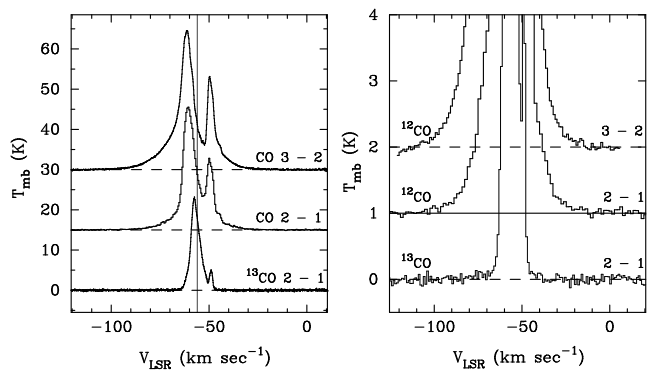}

\figcaption[]{
\label{fig-co-jcmt}
Left panel: Long integration single dish (JCMT) spectra of  CO \jtra21,
\jtra32, and $^{13}$CO \jtra21. The red-shifted emission peak at $\sim$
$-$49 km~s$^{-1}$ is due to an extended cloud unrelated to the outflow.
The systemic velocity at $-$56 km~s$^{-1}$ is marked with a vertical
ine. Right panel: The same spectra binned over 6 or 8 channels and
plotted to show the faint high velocity emission.}
\end{figure}

\begin{figure}
\includegraphics[ scale=0.7]{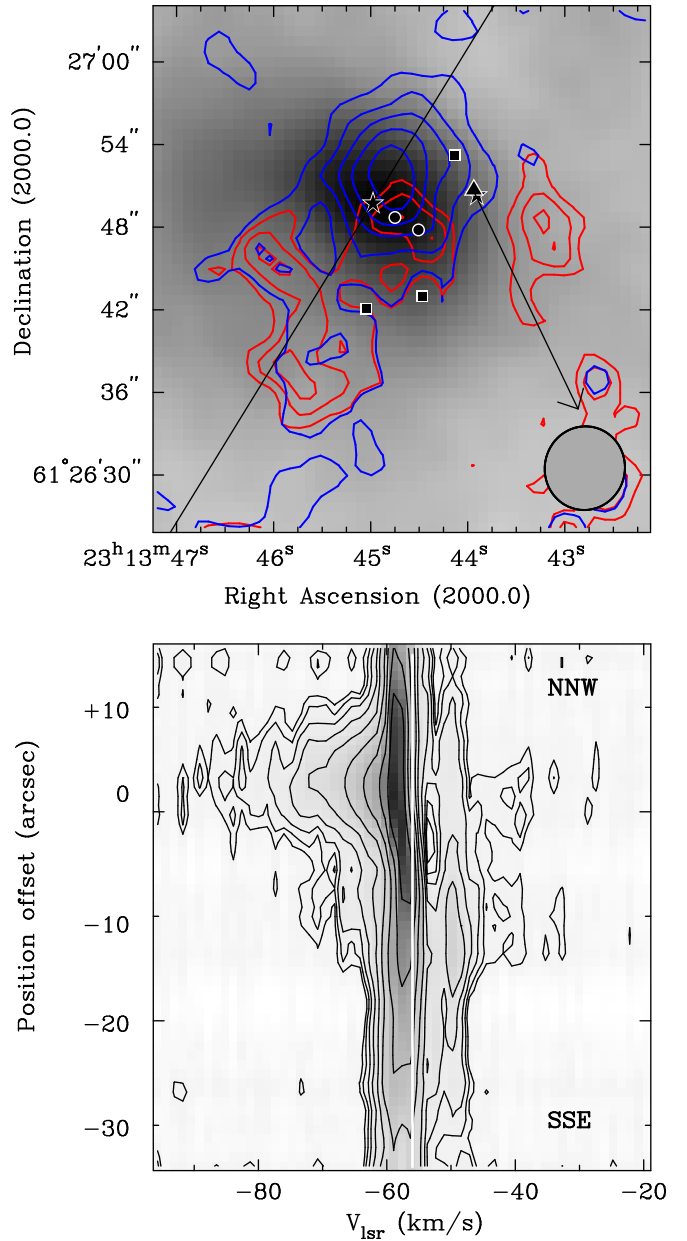}

\figcaption[]{
\label{fig-hco+outflow}
Top: Contour plot of the HCO$^+$ outflow overlaid on the integrated
H$^{13}$CN emission (grey scale), which shows the distribution of the
dense gas in the NGC\,7538\,S cloud core. The blue-shifted high velocity
emission is integrated from  $-$61 km~s$^{-1}$ to $-$80.7 km~s$^{-1}$
and plotted with blue contours. The contour levels are linear starting
at 5 K~km~s$^{-1}$ with a step size of 5 K~km~s$^{-1}$. The red-shifted
high velocity emission is integrated from  $-$49 km~s$^{-1}$ to $-$38.5
km~s$^{-1}$ and plotted with red contours. The contour levels are linear
starting at 4 K~km~s$^{-1}$ with a step size of 2 K~km~s$^{-1}$. The
position of NGC\,7538\,S and IRS\,11\,S are marked by star symbols while
the two new CARMA sources are marked with filled circles. The filled
squares are Class I methanol masers and the triangle is an H$_2$O maser
coinciding with  IRS\,11\,S . The HPBW is indicated in the bottom right
corner. The black line through NGC\,7538\,S shows the outflow axis at an
position angle of 148\degr. The red-shifted peak at 12\arcsec\ west of
NGC\,7538\,S is due to another outflow, possibly powered by IRS\,11\,S.
It is also seen in SiO (Figure \ref{fig-sio-outflow}{}). Bottom:
Position velocity plot as grey scale with contours along the outflow
axis. The contour levels are logarithmic and plotted with 10 contours
starting at 0.25~K to 14~K. The systemic velocity is marked by the white
vertical line.}
\end{figure}

\subsubsection{Outflow morphology}
\label{Outflow_morphology}

For NGC\,7538\,S we have many data sets that probe the outflow. Here we
will largely restrict the discussion to CO \jtra21, CO \jtra32, HCO$^+$
\jtra10 and SiO \jtra21. CO is a very good outflow tracer and chemically
much more robust than HCO$^+$, although at near cloud velocities the
emission can be  optically thick. Figure \ref{fig-co-jcmt} shows long
integration spectra  of CO \jtra21, \jtra32, and $^{13}$CO \jtra21, all
obtained with JCMT.  What we immediately see from these spectra is that
the outflow is an extreme high-velocity outflow \citep{Choi93}, because
we see high-velocity wings extending to velocities $\geq$ 55 km~s$^{-1}$
in both the blue- and the red-shifted side of the CO \jtra21 and \jtra32
spectra. The emission is strongly skewed to blue-shifted velocities,
because of strong self-absorption. The self-absorption is even more
extreme in HCO$^+$ \jtra10, where the red-shifted wing is almost
completely absent (Figure \ref{fig-spectra}) due to the strong accretion
flow, see Section \ref{Accretion_rate}. The strong emission peak at
$-$49 km~s$^{-1}$, which dominates the emission on the red-shifted side
of the CO lines, is due to the extended ``$-$49 km~s$^{-1}$'' molecular
cloud component,  which is seen over most of the NGC7538 molecular
cloud. The strong red-shifted self-absorption and the ``$-$49
km~s$^{-1}$'' cloud make it difficult to derive accurate parameters for
the red-shifted {\bf lobe} and we will therefore restrict most of our
discussion to the blue-shifted lobe.

\begin{figure*}[t]
\includegraphics[ scale=1,angle=-90]{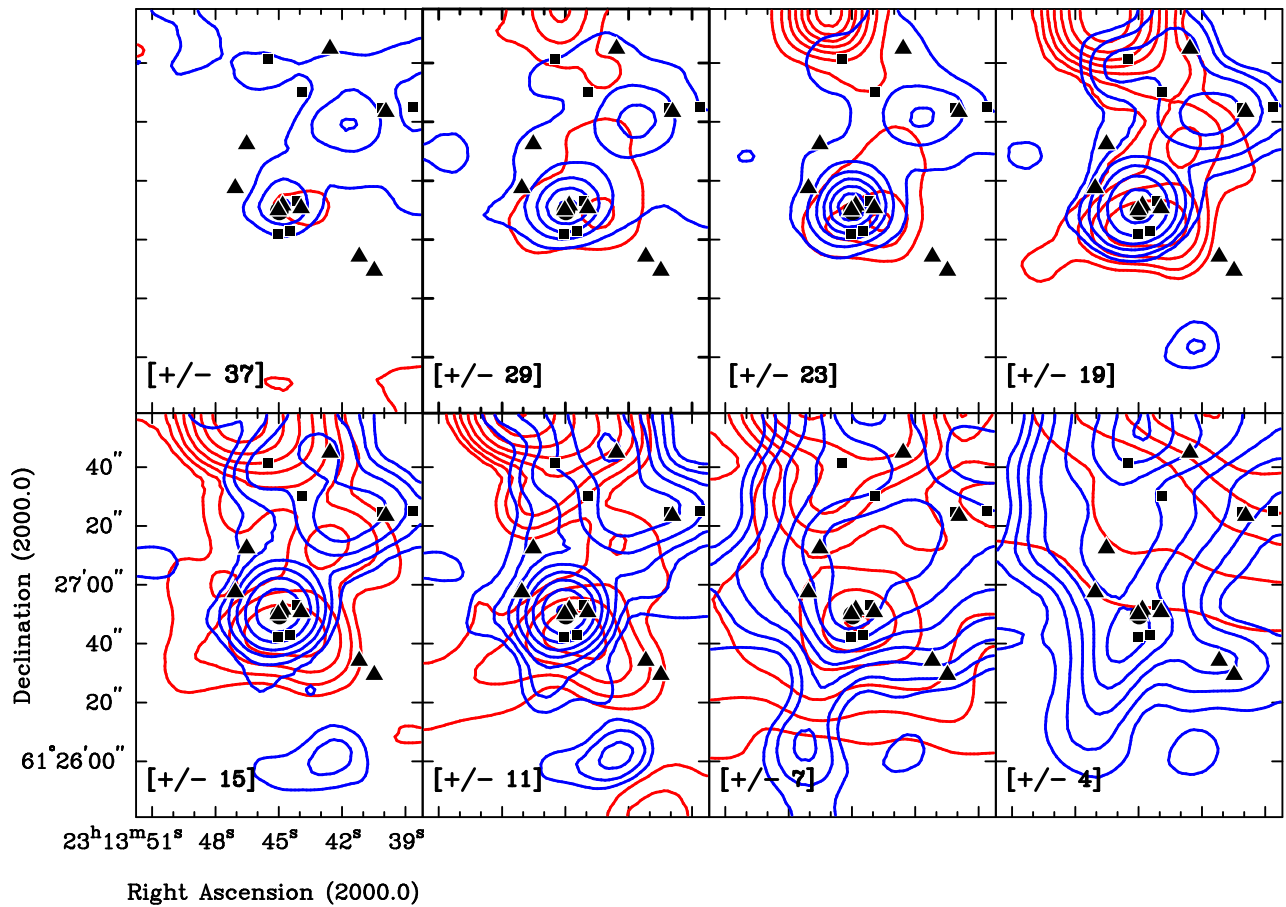}

\figcaption[]{
\label{fig-co32-outflow}
Maps of CO \jtra32 high velocity emission integrated over eight
different velocity intervals in blue- and red-shifted emission.  The
first velocity interval, bottom right panel, is 2 km~s$^{-1}$ centered
at a velocity $\pm$ 4 km~s$^{-1}$, the next five are integrated over 4
km~s$^{-1}$ wide velocity windows, and the last two over 8 km~s$^{-1}$
wide velocity intervals. The offset from the systemic velocity is
indicated at the bottom right of each panel. The HPBW for these
observations is $\sim$ 14\arcsec, which  is insufficient to resolve the
outflow(s) from the NGC\,7538\,S core. At the very north we can see the
strong red-shifted outflow from NGC\,7538\,IRS\,1 (outside the area
shown in these maps) and $\sim$ 40\arcsec\ northwest  of NGC\,7538\,S,
we see the blue-shifted lobe of another outflow.  The contour levels
start at 0.75 K~km~s$^{-1}$ with a step of 7.5 K~km~s$^{-1}$. The filled
triangles mark known H$_2$O masers, the squares CH$_3$OH Class I masers
and the filled circle the position of NGC\,7538\,S.
}
\end{figure*}

\begin{figure}[b]
\includegraphics[ scale=0.45,angle=-90]{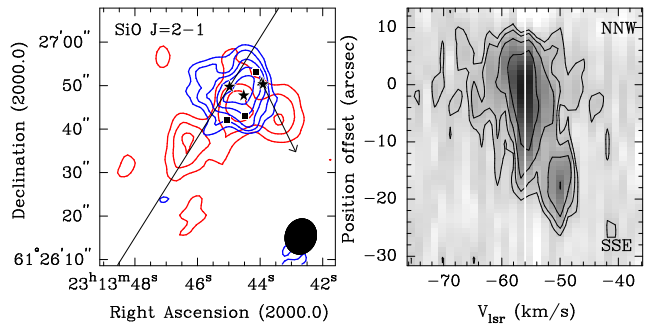}

\figcaption[]{
\label{fig-sio-outflow}
Left panel: Contour plot of the integrated SiO \jtra21 high velocity
emission. The blue-shifted high velocity emission is integrated from
$-$61.4 km~s$^{-1}$ to $-$70.9 km~s$^{-1}$ and plotted with blue
contours. The contour levels are linear starting at 1 K~km~s$^{-1}$ with
a step of 1 K~km~s$^{-1}$. The red-shifted high-velocity emission is
integrated from $-$52 km~s$^{-1}$ to $-$39.8 km~s$^{-1}$ and plotted
with  red contours. The contour levels are linear starting at 2
K~km~s$^{-1}$ with a step of 1 K~km~s$^{-1}$.  The black line shows the
position angle (148\degr{}) of the outflow, as determined from HCO$^+$
\jtra10. Labeling is the same as in Figure \ref{fig-hco+outflow}. Right
panel: Position velocity plot as grey scale with contours along the
outflow axis. The contour levels are linear starting at 0.3 K with a
step of 0.6 K. The systemic velocity is marked by the white vertical
line.
 }
\end{figure}

Figure \ref{fig-hco+outflow} shows the outflow imaged in HCO$^+$. The
outflow is clearly bipolar and very  compact, suggesting that we are
seeing a very young outflow. The position angle (p. a.) of the outflow
determined from the HCO$^+$ high velocity emission is $\sim$ 148\degr\
$\pm$ 4\degr, i.e. roughly perpendicular to the position angle we derive
for the accretion disk from H$^{13}$CN \jtra10 (Section \ref{h13cn}).
The position angle of the outflow agrees with the position angle of the
thermal jet (Paper II, see also Figure~\ref{fig-1mm-dcn}), 145\degr\
$\pm$ 2\degr, which is to be expected if the outflow is powered by the
jet. The blue-shifted  lobe is extremely compact and extends only $\sim$
8\arcsec\ to the NNW. The counter flow to the SSW contains both blue-and
red-shifted high velocity gas, suggesting that the red-shifted lobe is
somehow closer to the plane of the sky. The red-shifted lobe is more
extended and less collimated. It looks more like a half shell without a
clear front, suggesting that it has cleared the dense core and is
expanding into a lower density regime. The extent of the red HCO$^+$ 
lobe is $\sim$ 16 - 17\arcsec\ and brighter on the western side.
Examination of our CO \jtra32 map (Figure \ref{fig-co32-outflow}) shows
a similar size in CO. The SiO \jtra21 outflow,  however, appears more
jet-like and extends to $\sim$22\arcsec\ from the central protostar
(Figure \ref{fig-sio-outflow}). 

HCO$^+$ and SiO show a red-shifted lobe of another outflow west of
NGC\,7538\,S.  It is also seen in our CO \jtra32 map, especially at high
red-shifted velocities.  It does not stand out very clearly in our
single dish corrected HCO$^+$ image (Figure \ref{fig-hco+outflow}),
which shows a strong broad red-shifted emission peak at $\sim$
10\arcsec\ west of NGC\,7538\,S.  Inspection of the HCO$^+$ data sets
and the SiO image, however, suggests that it has a p.a. of  $\sim$
205\degr\ and that it is brighter on the western side in HCO$^+$. In
Figure  \ref{fig-hco+outflow} this outflow is shown with an arrow
outlining the origin,  extent ($\geq$ 17\arcsec{}),  and p.a. of the
outflow.  This red-shifted outflow has a rather high outflow velocity,
$\sim$ 25 - 30 km~s$^{-1}$, in  both HCO$^+$ and SiO. It is almost
certainly powered by the bright  IRAC and H$_2$O maser source,
IRS\,11\,S \citep[Paper II]{Corder08}. There is no clear sign of a
blue-shifted counter flow, although such a flow would overlap with the
blue outflow lobe from NGC\,7538\,S.

\subsubsection{Physical properties of the outflow}

The high velocity emission is  much stronger in CO \jtra32 than in
\jtra21 (Figure \ref{fig-co-jcmt}), which indicates  that the outflow is
hot. Since the beam size is smaller for the CO \jtra32 than the \jtra21
transition, the higher intensities in CO  \jtra32 than in \jtra21 are
partly due to better coupling to the beam, but detailed analysis shows
that the observed intensity ratios require the high velocity emission to
have temperatures of 100 K or more. We do have maps  of both CO J = $2
\to 1$ and CO  J = $3 \to 2$ and we do know the size of the blue outflow
lobe, enabling us to correct for beam filling. Nevertheless, it is
difficult to derive an accurate temperature from only two CO
transitions. We therefore prefer to use the ratio of the two
formaldehyde transitions, H$_2$CO,  J =  3$_{03}  \to 2_{02}$ and
3$_{22}  \to 2_{21}$, which were observed simultaneously with both BIMA
and JCMT. These observation are therefore insensitive to calibration and
pointing errors, and even though the line wings are much fainter in
H$_2$CO than in CO  we get a more accurate determination of the gas
temperature in the outflow. The ratio T$_b$($3_{03} -
2_{02}$)/T$_b$($3_{22} - 2_{21}$) is very sensitive to temperature,
especially if the gas is not too hot \citep{Mangum93}. If we integrate
the blue-shifted wing from $-$59 km~s$^{-1}$ to $-$63 km~s$^{-1}$, we
find a ratio of 2.65 at the position of the protostar and $\sim$ 2.9 in
the blue outflow lobe. These line ratios correspond to kinetic
temperatures of $\sim$ 120 K at the position of the protostar and $\sim$
100 K in the outflow lobe \citep{Mangum93}. For the red-shifted outflow
lobe integrated over the velocity range $-$53 km~s$^{-1}$ to $-$50.5
km~s$^{-1}$ the  ratios are 2.5 and 3.2, i.e. 140 K - 80 K, at the
position of the protostar and at the peak of the outflow lobe,
respectively. The H$_2$CO  data therefore confirm what we already saw
from CO, i.e. the temperature of the gas in the outflow $\geq$ 100 K.  A
population diagram analysis of methylcyanide, CH$_3$CN \jtra{12}{11},
which also traces the high velocity gas (see Section \ref{mcn}), gives a
rotational temperature of 147 $\pm$ 40 K, in good agreement with what we
derive from CO and H$_2$CO. Our H$_2$CO data suggest that the high
velocity gas in the blue outflow is hotter near the surface of the disk
and somewhat colder in the outflow lobe. The data also suggest that the
blue-shifted gas could be somewhat hotter in the blue-shifted gas than
the red-shifted gas, but this should be viewed with caution, since the
red-shifted emission is affected by self-absorption from the infalling
envelope.

Since the  HCO$^+$ is often enhanced in outflows, it is not a good mass
tracer \citep{Sandell05}. We therefore prefer to use $^{12}$CO, which is
more chemically robust. We have maps of the whole NGC\,7538\,S region in
$^{12}$CO \jtra21, i.e. the map published by \citet{Davis98}, $^{13}$CO
\jtra21, and  in $^{12}$CO \jtra32. We also have long integration
spectra  towards NGC\,7538\,S in  $^{12}$CO \jtra21, $^{13}$CO \jtra21,
and $^{12}$CO \jtra32 (Figure \ref{fig-co-jcmt}). Even though the 
$^{12}$CO \jtra32 goes deeper and has better spatial resolution than the
$^{12}$CO \jtra21 map, we will use the latter, because we  have only a
$^{13}$CO map in \jtra21, which allows  us to estimate the optical depth
in the outflow. The long integration  $^{13}$CO spectrum shows high
velocity wings over the velocity range $-$70 km~s$^{-1}$ (possible even
$-$80 km~s$^{-1}$) to $-$38 km~s$^{-1}$, confirming that most of 
$^{12}$CO emission is optically thick. The main disadvantage of using
single dish $^{12}$CO data,  is that the CO emission is very strong and
extended over a large velocity range, from about $-$65 km~s$^{-1}$ to
$-$48 km~s$^{-1}$, even though the blue outflow lobe can be seen in
channel maps at lower velocities (Figure \ref{fig-co32-outflow}). In
order to capture the low velocity gas we have therefore subtracted out
the extended background emission from the compact outflow emission. The
background emission was estimated in several areas and normalized to the
same area we integrated over in the map. For analysis of the long
integration spectra, we average spectra in the same background regions
and integrated the background spectrum over the same velocity range. 
Since the $^{12}$CO \jtra21 map has relatively poor signal-to-noise, we
found that we get more accurate mass estimates by using the long
integration CO spectra rather than integrating over the map, since the
outflow is very compact and we know the size very well from our BIMA
imaging in HCO$^+$. Furthermore, using long-integration CO spectra
allows us to trace the outflow to much higher velocities, which is
important for estimating the energy of the outflow. Because we had to
create reference spectra in both $^{12}$CO and $^{13}$CO from positions
outside the outflow, the biggest uncertainty in our mass estimate is due
to the opacity correction at low velocities, which depends on how well
we subtracted out the extended cloud emission. If we assume that the
bulk of the gas has a temperature of 100~K, and a [CO/H$_2$] =
10$^{-4}$, our $^{12}$CO \jtra21 data give an outflow mass of  6.5
\Msun, over the velocity range  $-$60 km~s$^{-1}$ to $-$111 km~s$^{-1}$.
About 75\% of this mass resides at low outflow velocities, $<$10
km~s$^{-1}$. The  momentum of the blue outflow lobe, P$_{blue}$ $\sim$
50 ${\rm M_{\odot} ~km ~s^{-1}}$, and since it is natural to assume that
the  momentum is conserved, we get P $\sim$ 100 ${\rm M_{\odot} ~km
~s^{-1}}$ for the whole outflow. These values are not corrected for the
inclination of the outflow. We should note, however, that the blue
outflow lobe most likely also includes the outflow from other sources in
NGC\,7538\,S (see paper II), and perhaps even the IRS\,11\,S outflow,
see Section \ref{Outflow_morphology}.

Since we now have mass estimates of the outflow  from CO, we can
determine how much the HCO$^+$ abundance is enhanced in the outflow
using our combined BIMA and FCRAO outflow map.  If we assume normal
HCO$^+$ abundances, i.e. [HCO$^+$/H$_2$] = 2.8~10$^{-9}$, and optically
thin HCO$^+$ emission \citep{Hogerheijde98}, we get an outflow mass of 
31.6 \Msun, suggesting that HCO$^+$ is enhanced by about a factor of
five. This enhancement is more modest than in the IRS\,9 outflow, where
\citet{Sandell05} found HCO$^+$ enhanced by a factor of 30. However, if
we only compare the mass estimates at high velocities, we find that the
HCO$^+$ abundance is enhanced by a factor of  20 - 40, suggesting that
enhancement of HCO$^+$ is caused by shocks from the high velocity
outflow. 

In order to determine the mass loss rate and energetics of the outflow,
we need to know the age of the outflow. The age is typically determined
from the dynamical time scale, corrected for the inclination of the
outflow \citep{Cabrit90}.  It is difficult to accurately estimate the
inclination of the outflow. However, since we have a wide-angled outflow
and there is some overlap between the blue and the red-shifted outflow
lobe, a comparison of the different model cases computed by
\citet{Cabrit90}, suggests an inclination angle between 40\degr\ -
60\degr.  In the following we assume an inclination angle of 50\degr. If
we take the mass weighted outflow velocity for the blue outflow lobe,
which is $<$V$>$ = 7.3~km~s$^{-1}$,  we get  dynamical time scale, t$_d$
= 12,200 yr. This is likely to be an under-estimate, since it does not
account for the time it has taken the outflow to expand into the
massive, dense cloud core surrounding the outflow. We will therefore
also examine the red outflow lobe.  Here we cannot derive a reliable
mass estimate, because the red-shifted gas  velocity gas is heavily
self-absorbed, especially at low velocities. We therefore use the
terminal velocity at the tip of the flow, $\sim$ 0.3 pc from the star
(22\arcsec{}), which is 17~km~s$^{-1}$ for HCO$^+$ \jtra10. In this case
we get a dynamical time,  t$_d$ =  14,000 yr,  about the same as we
derived for the blue outflow lobe. It is probably an under-estimate,
because the mass-weighted velocity is likely to be lower than the
terminal velocity that we used. We can get a third estimate of the
outflow rate from theoretical models of collapsing cloud cores, which
suggest that outflows form at about 0.6 times the free-fall time,
$\tau_{ff}$ \citep{Tomisaka98}. The average gas density in the
elliptical core surrounding NGC\,7538\,S is $\sim$ 10$^7$ cm$^{-3}$,
resulting in $\tau_{ff}$  = 2.5~10$^4$ yr, which would suggest that the
outflow started about 15,000 yrs ago, in reasonable agreement with what
we deduced from the dynamical time scales. If we assume an outflow age
of 15,000 yr, we find a mass loss rate of $\sim$  4.3 10$^{-4}$ ${\rm
M_{\odot} ~yr^{-1}}$, if  the red outflow lobe has a similar mass as the
blue lobe. With these assumptions we derive a momentum flux, F =  1.0
10$^{-2}$ ${\rm M_{\odot}~km~s^{-1}~yr^{-1}}$,  a mechanical luminosity
of  14.2 \Lsun, and a kinetic energy of  1.3 10$^3$ ${\rm
M_{\odot}~(km~s^{-1})^2}$ or 2.6 10$^{46}$ ergs. The derived momentum
flux would suggest  that the exciting star or stars have a bolometric
luminosity of 10$^4$  \Lsun, if we use the correlation by
\citet{Beuther02a} derived for Class I and young high-mass objects,
which is close to the observed luminosity.  For Class 0 objects
\citeauthor{Bontemps96} found an an outflow efficiency  $\sim$ 10 higher
than for Class I objects, but this is not true for NGC\,7538\,S, even
though  in any other respect it appears to be a Class 0 object. It is
possible that this relationship does not hold for extremely young
outflows, which are still density bounded.

\subsubsection{Accretion flow}
\label{Accretion_rate}

\begin{figure}[htbp]
\centering
\begin{minipage}[b]{10cm}
\includegraphics[ width=8.5cm]{f8a.eps}
\end{minipage}
\begin{minipage}[b]{8.7cm}
\includegraphics[ width=8.8cm]{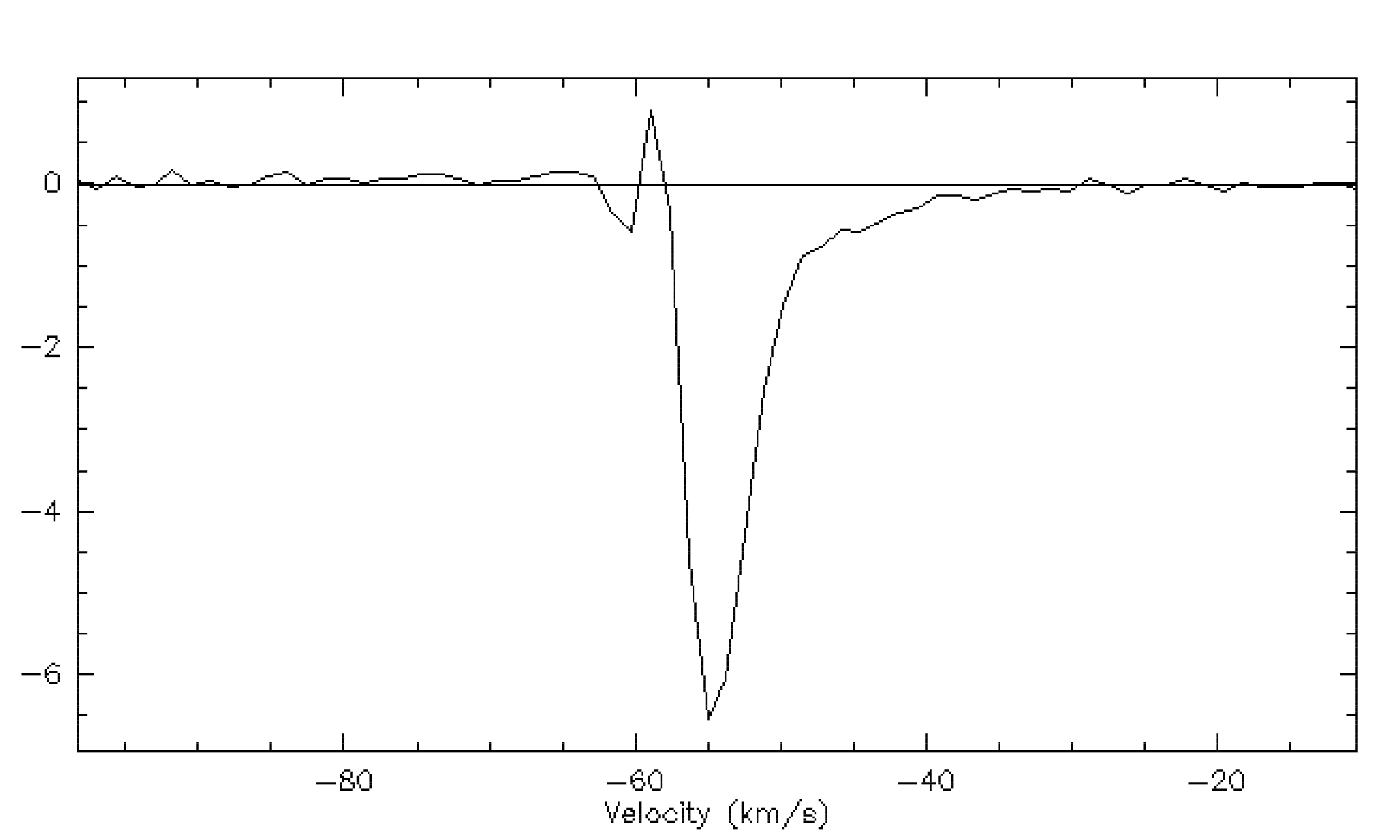}
\end{minipage}

\figcaption[]{
\label{fig-accretion-profiles}
The top panel shows the HCO$^+$ \jtra10  spectrum towards NGC\,7538\,S.
Overlaid on the HCO$^+$ profile, we have plotted the SO J = $5_5  \to 
4_4$  line at the same position in red, which has been scaled to match
HCO$^+$ spectrum. This SO transition is not affected by self-absorbtion
and shows symmetric blue-and red-shifted wings from the outflow.  In the
bottom panel we show the residual of a two component Gaussian fit to the
blue side of the HCO+ spectrum. This residual represent the predicted
absorption profile of cold optically thick gas in front of the hot 
NGC\,7538\,S outflow assuming that in absence of absorption, the HCO$^+$ 
profile would be similar to SO, i.e. have symmetric blue-and red-shifted
wings. The absorption profile is dominantly red and extends to $\sim
-$40 km~s$^{-1}$.
}
\end{figure}

 The red-shifted outflow wing is almost completely absent in HCO$^+$
 \jtra10 (Figure \ref{fig-spectra}),  suggesting that it is
 self-absorbed by the cold, infalling cloud envelope. We know that
 NGC\,7538\.S is surrounded by a dense, cold ($\sim$ 25 K) massive
 envelope ($\sim$ 2,000\Msun{}), where most molecules (including
 HCO$^+$) are optically thick (Section \ref{core}). If the cloud core
 was in thermal equilibrium, one would therefore expect to see a
 symmetric line profile around the systemic velocity of the cloud core,
 but the absorption appears to affect only the red-shifted velocities.
 Our long integration CO spectra, Figure \ref{fig-co-jcmt}, see Section
 \ref{Outflow_morphology}, show that the outflow velocities are the same
 in the blue- and the red-shifted wing. Therefore the absence of
 red-shifted HCO$^+$ emission is not because the red-shifted emission is
 absent, but because it is self-absorbed by the outflow. This is
 confirmed by our observations of  SiO \jtra21, and SO  J = $5_5  \to
 4_4$, which both show symmetric high velocity wings over the velocity
 interval, where emission from HCO$^+$ is very weak or absent. Since the
 SiO and SO transitions come from much higher energy levels, the absence
 of self-absorption in these lines suggest that they are not
 sufficiently excited in the cold surrounding cloud envelope to cause
 self-absorption against the hot outflow. The red-shifted
 self-absorption, which is so striking in HCO$^+$ must therefore be
 caused by infall of the cloud core surrounding NGC\,7538\,S. To
 illustrate this more clearly, we have taken the SO J = $5_5  \to  4_4$
 line, which was observed with a similar beam size as HCO$^+$ and scaled
 it, so that it roughly matches the HCO$^+$ profile (Figure
 \ref{fig-accretion-profiles}). From this figure we can see that infall
 starts to affect the red-shifted outflow from velocities $\sim$ $-$40
 km~s$^{-1}$. We can directly create an infall profile by fitting two
 Gaussians (one for the cloud core and one for the outflow) to the
 blue-shifted side of the HCO$^+$ profile and blanking out the part of
 the spectrum which is likely to be affected by self-absorption. If we
 subtract the fit from the observed spectrum, we get the absorption
 profile shown in Figure~\ref{fig-accretion-profiles}, which confirms
 that we can see the infall to at least 16 km~s$^{-1}$. At the systemic
 velocity, HCO$^+$ is optically thick. The ratio of the the peak
 intensity of HCO$^+$ to H$^{13}$CO$^+$  (Table \ref{tbl-3},
 Figure~\ref{fig-accretion-profiles}) is $\sim$ 2, if we use the fitted
 peak for HCO$^+$, 6.5 K, and ignore any differences in beam size. This
 corresponds to an optical depth of $\sim$ 20 (for a [$^{12}$C/$^{13}$C]
 abundance of 40). At the peak of the absorption profile we therefore
 have an optical depth of 10. Inspection of position velocity plots
 (Figure \ref{fig-hco+outflow}) and  spectra around NGC\,7538\,S
 indicates that we see narrow self-absorption out to a radius of at
 least 15\arcsec\, while the self-absorption at high velocities  is only
 seen in the immediate vicinity of NGC\,7538\,S, indicating  that the
 accretion flow is accelerating. In order to estimate the mass of the
 infalling cloud envelope, we somewhat arbitrarily divide the absorption
 profile into two parts: an extended low velocity component (V $>$ $-$
 54 km~s$^{-1}$) with a FWHM of $\sim$ 6\arcsec, and an unresolved
 component extending to velocities $\sim$ $-$40 km~s$^{-1}$. The low
 velocity infall component has an integrated intensity of $\sim$ 14
 K~km~s$^{-1}$ and an assumed average optical depth of $\sim$ 3, while
 the integrated intensity of the unresolved ``high velocity'' infall is
 23.5 K~km~s$^{-1}$. Since we see the gas in absorption, it must have an
 optical depth about one or more. If we sum up the gas seen in
 absorption, we get $\sim$ 125 K~km~s$^{-1}$, which is half the
 infalling gas, since we have an equal amount from the backside of the
 cloud. If  we adopt an [HCO$^+$/H$_2$] abundance of 2.8 10$^{-9}$ and a
 kinetic temperature of 25 K, we get a mass of 21 \Msun, which is
 strictly speaking a lower limit. If we assume that the  time  scale for
 accretion is the same as for the outflow, i.e. 15,000 yr, we get a mass
 accretion rate $\sim$ 1.4 10$^{-3}$ M$_{\odot}$~yr$^{-1}$.
 
 A more common way to estimate the accretion rate is to assume that the
 outflow rate is proportional to the accretion  rate, see e.g.
 \citet{Beuther02a}. If we assume that the ratio of the outflow rate and
 the accretion rate is approximately 0.3 \citep{Tomisaka98,Shu99} we get
 an accretion rate of 1.4 10$^{-3}$ M$_{\odot}$~yr$^{-1}$, the same as
 what we got from our direct estimate of the infalling cloud mass. The
 agreement is probably fortuitous, because both estimates have an
 uncertainty of at least a factor of two or three.

 Another way to estimate the accretion rate is to assume that we have a
 steady state accretion flow with an infall velocity, v$_{inf}$, onto a
 uniform disk with a radius, R, and a surface density, $\Sigma$, see
 e.g.  \citet{Beltran04}. In this case the mass accretion rate,
 $\dot{\rm M}_{acc}$, can be expressed as:  $\dot{\rm M}_{acc}$ =
 2$\pi\Sigma$Rv$_{inf}$.  \citet{Allen03}, who did numerical simulations
 of the collapse of a  magnetized molecular cloud core assumed to
 undergo rigid rotation, show that in such a case the infall velocity is
 the same as the rotation velocity of the disk.  Since we measure a
 rotational velocity of $\sim$ 1.3 km~s$^{-1}$ at a radius of  $\sim$
 7000 AU (Section \ref{rotating_disk}) and with a surface density,
 $\Sigma$ = 2.1 g~cm$^{-2}$, deduced from our H$^{13}$CN observations of
 the disk (Section \ref{compact_core}), we find an accretion rate,
 $\dot{\rm M}_{acc}$ = 2.8 10$^{-3}$ M$_{\odot}$~yr$^{-1}$, twice as
 large than our two other estimates, but well within the errors of each
 method.

\subsection{The compact core and the accretion disk}
\label{compact_core}

\begin{figure}
\includegraphics[ scale=0.38,angle=-90]{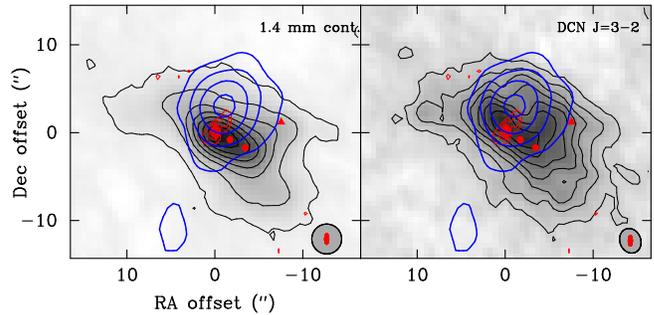}
\figcaption[]{
\label{fig-1mm-dcn}
Gray scale images of  total integrated DCN \jtra32 emission (right
panel) and 1.4~mm continuum emission(left panel) overlaid with black
contours. Overlays: 3.6 cm continuum emission (red contours), and HCO+
blue shifted emission (blue contours). The red triangle marks the H$_2$O
maser associated with the mid-IR source IRS\,11\,S and the filled
circles the two new CARMA sources discovered in the elliptical core.
}
\end{figure}

\begin{figure}[t]
\includegraphics[ scale=0.35,angle=-90]{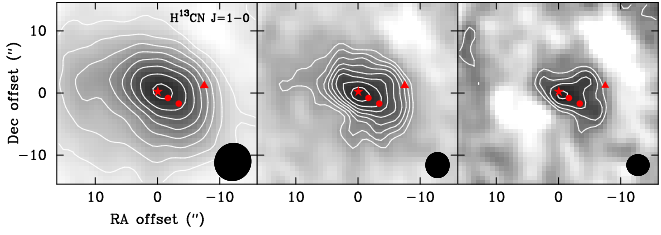}
\figcaption[]{
\label{fig-h13cndisk}
Grayscale images of Integrated H$^{13}$CN \jtra10 emission overlaid with
contours for three different angular resolutions. The image to the left
is created from all data sets using natural weighting, while the one in
the middle the has been made with robust weighting and excluding all
uv-spacings shorter than 5 k$\lambda$. The image to the right, which has
the highest angular resolution, was made with natural weighting and
excluding all uv-spacings shorter than 10 k$\lambda$, shows the disk
most clearly, but has rather poor signal-to-noise. The symbols are the
same as in Fig.~\ref{fig-1mm-dcn}, except for NGC\,7538\,S, which is now
marked with a red star symbol. }
\end{figure}

Both DCN \jtra32 and H$^{13}$CN \jtra10 show an elliptical source
approximately centered on NGC\,7538\,S superposed on more extended
fainter emission. In the left panel of Figure \ref{fig-1mm-dcn} we show
the integrated DCN emission, overlaid with the 3.6~cm continuum emission
from  Paper II, and the blue-shifted HCO$^+$ emission, which show how
the jet and the outflow relate to the disk-like structure seen in the
integrated DCN image. In the right panel of Figure  \ref{fig-1mm-dcn} we
show the 1.4~mm continuum emission (Paper II), also overlaid with the
blue-shifted HCO$^+$ emission. The dust continuum and the DCN emission
look very similar, although there appears to be more structure in the
DCN emission than in the dust continuum emission, even though both have
been imaged with very similar spatial resolution. The integrated DCN
emission shows a ridge-like structure extending to the northwest $\sim$
4\ptsec5  west of NGC\,7538\,S. This ridge or DCN ``clump'' is not seen
in the dust continuum, but it does show up in H$^{13}$CN, but only at
high spatial resolution (Figure  \ref{fig-h13cndisk}, right panel). As
we can see from Figure \ref{fig-1mm-dcn} this DCN condensation borders
the western side of the blue outflow, and could therefore be a dense gas
clump compressed by  the outflow (Section ~\ref{dcn}). A few arcseconds
west of the DCN clump is another IRAC (mid-infrared) source, IRS\,11\,S,
but the DCN clump is rather compact ($\sim$ 3\arcsec{}) and the
positional offset is too large to suggest an association with this
source. IRS\,11\,S is is clearly another young star, because it excites
an H$_2$O maser and powers a small outflow.

Gaussian fits to the dust continuum and to the integrated H$^{13}$CN
emission give a size for the compact core of $\sim$ 10\arcsec\ $\times$
4\arcsec\ at p.a. 58\degr\ $\pm$ 3\degr, centered $\sim$ 2\arcsec\
southwest of the center of the accretion disk. If we assume a gas
temperature of 35~K,  and an [H$^{13}$CN/H$_2$] abundance of 1.25
$\times$ 10$^{-10}$, i.e. the same abundance that we assumed for the
NGC\,7538\,S cloud core, see Section \ref{core}, we obtain a total mass
of 85\Msun. This mass estimate is very similar to what we derive from
our dust continuum observations, $\sim$125\Msun\ (Paper II). Even though
we do not have a good estimate for the gas temperature in the compact
core, our assumed temperature appears reasonable. The lowest measured
gas temperature comes from ammonia, 25 K. This gives a lower limit to
the gas temperature. The temperature of the disk is certainly $<$ 50 K 
(see Section~\ref{mcn}). Since the gas mass scales roughly linearly with
temperature, the largest error in our gas mass estimate is due to the 
uncertainty of the H$^{13}$CN abundance.

Of all the molecules that we have observed in the 3~mm band, it appears
that the best molecular tracer of the disk is H$^{13}$CN \jtra10. As far
as we can tell, the H$^{13}$CN emission is only weakly contaminated by
the high-velocity outflow, and we have sensitive maps obtained with
relatively high spatial spatial resolution ($\sim$ 3\ptsec9).  We also
observe a clear velocity gradient along the major axis of the disk in
HN$^{13}$C \jtra10 and H$^{13}$CO$^+$ \jtra10 in the same direction as
in H$^{13}$CN, i.e. blue-shifted emission to the northeast and
red-shifted emission to the southwest.   A position-velocity cut of
H$^{13}$CO$^+$ along the outflow direction shows blue-shifted emission
to the northwest and red-shifted emission to the southeast, i.e. 
H$^{13}$CO$^+$ is clearly affected by the outflow, which makes it very
difficult with this angular resolution to dis-entangle how much of the
observed velocity gradient comes from the disk and how much of it is due
to the outflow. For these lines we have only one B- and one C-array track
giving us an angular  resolution of $\sim$ 7\ptsec8 (Table \ref{tbl-1}).
Since  the data are rather noisy and have poorer spatial resolution than
H$^{13}$CN, we will not discuss them further in this paper.

At 1.4~mm we have high signal-to-noise maps in DCN \jtra32 and CH$_3$CN
\jtra{12}{11}, with even higher spatial resolution ($\sim$ 2\ptsec6)
than H$^{13}$CN \jtra10. However, although DCN \jtra32 appears to trace
the accretion disk, the emission is rather strongly affected by emission
from the outflow, making it difficult to accurately determine the
rotation curve for the disk, although the values we derive are in
reasonable agreement with  what we find from H$^{13}$CN (Section
\ref{h13cn}). Although \citet{Cesaroni99,Cesaroni05} find CH$_3$CN to be
a good disk tracer for the young high-mass star IRAS $20126+4104$, we
find that CH$_3$CN is  dominated by emission from the outflow, which
effectively masks any emission from the underlying accretion disk
(Section \ref{mcn}).

Below we discuss in detail the findings for these three molecules.

\begin{deluxetable}{lcc}
\tablecolumns{3}
\tablenum{4}
\tablecaption{Radial velocities and line-widths of H$^{13}$CN \jtra10  in the accretion disk}
\label{tbl-4}
\tablehead{
\colhead{Position\tablenotemark{a}} & \colhead{$V_{\rm LSR}$}&  \colhead{$\Delta V$}\\
\colhead{  }  &\colhead{ [km s$^{-1}$] }&\colhead{ [km s$^{-1}$] }\\
}

\startdata
\sidehead{HPBW = 6.0 $\times$ 5.7}
\cline{1-3}
northeast &   $-$56.59 $\pm$ 0.06  &  3.9   $\pm$ 0.1 \\
center      &   $-$55.98 $\pm$ 0.07  &  4.7  $\pm$ 0.1\\ 
southwest &  $-$55.19 $\pm$ 0.16  &  4.6   $\pm$ 0.4\tablenotemark{b}\\ 
\sidehead{HPBW = 4.0 $\times$ 3.7}
\cline{1-3}
northeast  &  $-$57.34 $\pm$ 0.11  &  3.1   $\pm$ 0.3 \\
center       & $-$55.97 $\pm$ 0.12  &  5.5  $\pm$ 0.1  \\
southwest &  $-$54.17  $\pm$ 0.59  &  4.4   $\pm$ 0.6\tablenotemark{b} \\
\sidehead{HPBW = 3.5 $\times$ 3.3}
\cline{1-3}
northeast  &  $-$57.76 $\pm$ 0.21  &  2.3   $\pm$ 1.0 \\
center\tablenotemark{c}        &  \nodata  &  \nodata \\
southwest & $-$54.33  $\pm$ 0.89  &  5.5   $\pm$ 0.8\tablenotemark{b} \\
\cline{1-3}
\enddata

\tablenotetext{a}{Positions southwest and northeast are at $\pm$2\ptsec5 from the center at a p.a. of 58\degr{}.}
\tablenotetext{b}{At the southwest position the red-shifted emission
from the disk is blended by emission from the surrounding cloud
core,  resulting in uncertain velocities  and broadened line widths.}
\tablenotetext{c}{The line is too deeply self-absorbed to result in a
successful fit.} 
\end{deluxetable}

\subsubsection{H$^{13}$CN \jtra10}
\label{h13cn}

\begin{figure}[t]
\includegraphics[ scale=0.34,angle=-90]{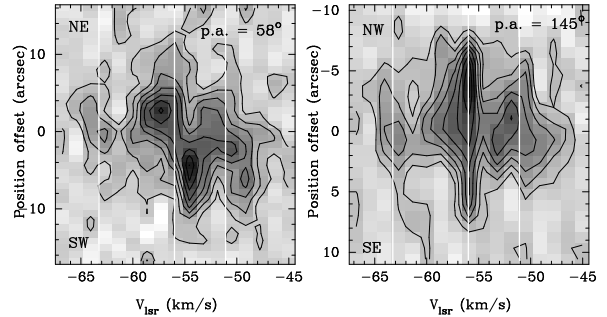}
\figcaption[]{
\label{fig-h13cncuts}
Position velocity cuts of H$^{13}$CN \jtra10 emission along the disk plane
(left panel) and along the outflow (right panel). The vertical white lines mark 
the 3 hyperfine components at the systemic velocity. }
\end{figure}

Emission from H$^{13}$CN \jtra10 is widespread over the whole NGC\,7538
molecular cloud, with the strongest emission in the whole cloud being
centered on NGC\,7538\,S. Here we are interested only in emission from
the disk, and we would prefer to completely filter out all the extended
emission surrounding the accretion disk. We have therefore created
several images, where we have excluded short spacings, and we
have also experimented with weighting the uv-data in different ways. 
However, since most of our data come from relatively short spacings,
this has a severe penalty in signal-to noise. Figure \ref{fig-h13cndisk}
shows the integrated H$^{13}$CN emission with three different angular
resolutions. We will base most of our analysis on the central image,
which was created by excluding all uv-spacings shorter than 5 k$\lambda$
and using robust weighting, giving an angular resolution of 
4\ptsec0 $\times$ 3\ptsec7,  p.a. = 0.2\degr, and still resulting in
a good signal-to-noise ratio.  Figure \ref{fig-h13cncuts} shows
position velocity plots along the disk-plane (p.a. = 58\degr{}) and in
the outflow direction (p.a. = 145\degr{}). The H$^{13}$CN emission shows
a very distinct velocity gradient over the disk, with blue-shifted
emission to the northeast and red-shifted emission to the southwest. At
the center of the accretion disk the main hyperfine component (F = $2
\to 1$) appears to be self-absorbed. The self-absorption is hardly
visible in spectra at lower angular resolution, i.e our naturally weighted map
(Figure \ref{fig-spectra}), while the line is almost completely absent at
our highest angular resolution. The highest blue- and red-shifted
emission peaks $\sim$ $\pm$ 2\ptsec5 from the center of the disk,
suggest a diameter of $\sim$ 5\arcsec\  or $\sim$ 14,000 A.U. for the
rotating disk. DCN \jtra32 gives similar results for the disk size (Section
~\ref{dcn}).

\begin{figure}[h]
\includegraphics[ scale=0.75,angle=-90]{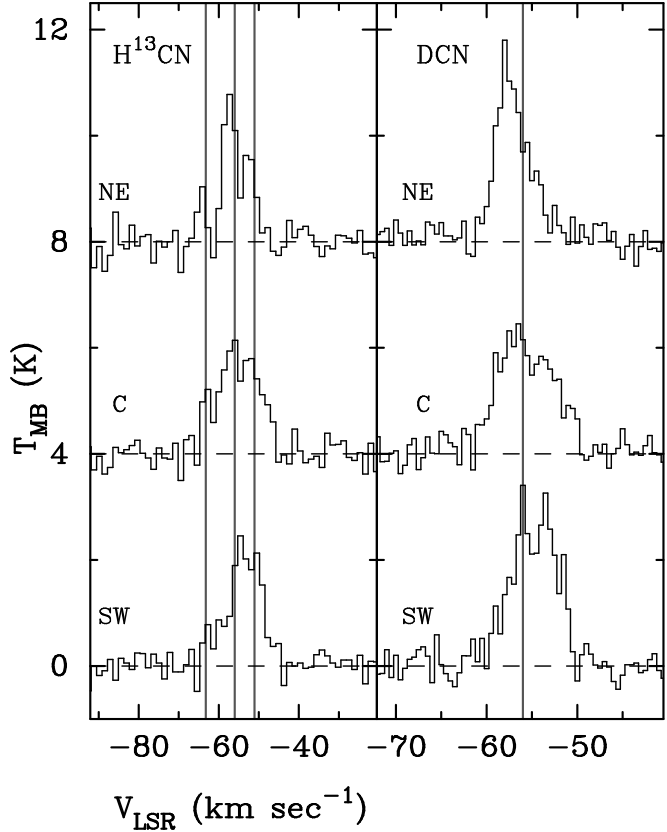}
\figcaption[]{
\label{fig-disk-spectra}
Spectra from the blue-shifted edge (NE, offset 2\farcs12,+1\farcs32),
center (0\arcsec,0\arcsec{}), and red-shifted edge (SW, offset
-2\farcs12,-1\farcs32 ) of the disk for  the high angular resolution,
robust weighted H$^{13}$CN \jtra21  image (left panel) and for DCN
\jtra32 (right panel).  For both molecules we have indicated the
systemic velocity and for H$^{13}$CN additionally the velocities of the
two hyperfine lines relative to the main line, F = $2 \to 1$.}
\end{figure}

We made Gaussian fits of spectra extracted from all three images in
order to more precisely quantify the velocity gradient. These Gaussian
fits were made using method hfs in CLASS, which results in very reliable
velocity information, since it uses the information from all three
hyperfine lines. We also made Gaussian fits by locking the hyperfine
lines to the main line, but since the lines are blended this require us
to keep the line width constant in order to get reliable fits. The two
methods give similar results to within $\pm$ 0.1 km~s$^{-1}$. In Table 
~\ref{tbl-4} we give the results obtained with method hfs. We can see
from this table that higher spatial resolution results in larger
velocity shifts at the edge of the disk and narrower line widths, as one
would expect, because we get less velocity smearing within the beam. At
the western side of the disk the emission from the surrounding cloud
core is blended with the emission from the disk. This is seen better in
DCN \jtra32, which has negligible hyperfine splitting, see Figure
\ref{fig-disk-spectra}.

As we can see from the position velocity plot along the outflow
(Figure~\ref{fig-h13cncuts}), even H$^{13}$CN shows some emission from the
outflow, because the emission northwest of the disk is blue-shifted,
while there may be some red-shifted emission to the southeast.

\subsubsection{DCN \jtra32}
\label{dcn}

\begin{figure}[t]
\includegraphics[ scale=0.6,angle=-90]{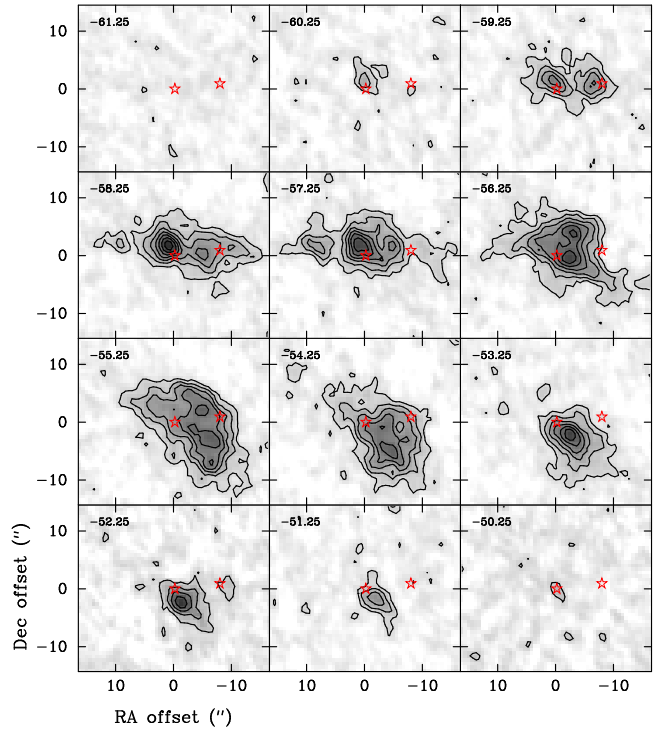}
\figcaption[]{
\label{fig-dcn-chann}
DCN \jtra32 channel maps in grayscale overlaid with contours. Each panel
shows the DCN emission integrated over  a 1 km~s$^{-1}$ wide velocity
interval. The center velocity of each map is indicated at the top left
of each panel. The two star symbols mark the position of NGC\,7538\,S
(at 0\arcsec,0\arcsec{}) and IRS\,11\,S. The contours are linear
starting at 0.12 Jy~beam$^{-1}$ with a step size of 0.12 Jy~beam$^{-1}$.
}
\end{figure}

The emission from DCN \jtra32 is more compact than H$^{13}$CN \jtra10
and also appears more lumpy than the H$^{13}$CN emission, c.f. Figures
\ref{fig-1mm-dcn} \& \ref{fig-h13cndisk}. To some extent this is to
be expected, since the spatial resolution is much higher in DCN than
in H$^{13}$CN (Tables \ref{tbl-1} \& \ref{tbl-2}). At the highest
angular resolution,  $\sim$ 3\ptsec5, that we could achieve from our
H$^{13}$CN observations  (Figure \ref{fig-h13cndisk}, right panel),
H$^{13}$CN  does appear to show a second condensation $\sim$ 4\ptsec5
west of NGC\,7538\,S, which is also seen in DCN. However, the difference
in  morphology between the DCN emission and H$^{13}$CN emission is too
large to be a result from angular resolution alone.  To get a better
idea where the DCN emission comes from, we  created a set of images
with 1 km~s$^{-1}$ spacing over the whole velocity range where we see DCN
emission. These channel maps are shown in Figure \ref{fig-dcn-chann}.
At the highest red- and blue-shifted velocities, the emission peaks on
or near the center of NGC\,7538\,S, while at near cloud velocities the
emission is stronger to the NE or the SW of the protostar, depending on
whether we look at blue-shifted or red-shifted emission. At slightly
blue-shifted velocities ($\sim$ $- $58~km~s$^{-1}$) DCN shows an E-W
ridge with three dominant peaks. The strongest one is slightly  north
northwest, $\sim$ 1 - 2\arcsec,  of  NGC\,7538\,S, with another fainter
core at +9\arcsec,+2\arcsec\ relative to NGC\,7538\,S. To the west there
is another peak at $-$5\arcsec,+1\arcsec\ and an even fainter one at
$-$10\arcsec,+1\arcsec. None of the DCN peaks, except the central one,
coincide with any known object in the NGC\,7538\,S cloud core. The H$_2$O
maser and IRAC source, IRS\,11\,S, the other known young active object
in the vicinity of NGC\,7538\,S, is roughly halfway between the two
western cores. At the cloud velocity the strongest emission is $\sim$
$-$2\arcsec\ west of S, but here we see another very strong peak at
$-$3\arcsec,+4\arcsec. The line at this position is centered at  the cloud
velocity  ($-$56 km~s$^{-1}$), and has a  very narrow line width, $\sim$ 1
km~s$^{-1}$, with a faint blue-shifted shoulder. At red-shifted velocities
most of the DCN emission comes from an extended  ($\sim$ 10\arcsec{})
region southwest of NGC\,7538\,S.  Although each velocity interval shows
several red-shifted DCN clumps, the position of these clumps move when
going from one velocity to the next, suggesting velocity gradients or
turbulent gas. This is also seen in spectra from the southwestern part of
the core, which are much broader with  line widths of 4 - 5 km~s$^{-1}$.

If we integrate over 5.5  km~s$^{-1}$ centered on V$_{lsr}$ =
$-$56  km~s$^{-1}$, i.e. roughly the velocity spread that we see in
H$^{13}$CN, we  see a central  ``disk-like''  feature, similar to what
we see in H$^{13}$CN, but with a northwestern ridge curving up from the
southwestern edge of the ``disk'' , i.e. very similar to the integrated
DCN emission shown in Figure \ref{fig-1mm-dcn}. However, the channel
maps show that this is not really  a ridge, but a blending of the  DCN
clump at $-$5\arcsec,+1\arcsec.  The double-peaked ``disk'', however, is
centered about somewhat west ($\sim$ 1\arcsec{}) of the protostar. Even
though this ``disk'' has roughly the same position angle as the continuum
and H$^{13}$CN disk, it is also shifted about one arcsecond to the north,
making it questionable whether it is really a disk.

\begin{figure}[t]
\includegraphics[ scale=0.375,angle=-90]{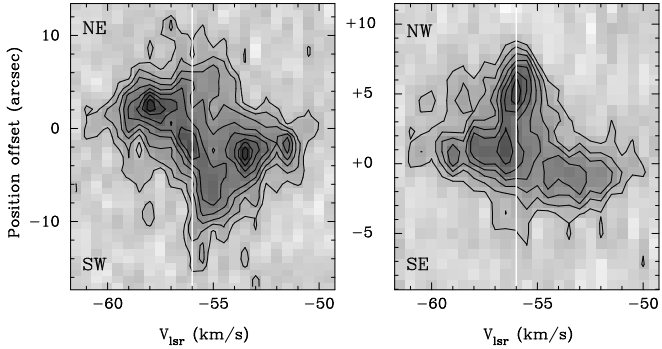}
\figcaption[]{
\label{fig-dcn-cuts}
Position velocity plots of DCN \jtra32 emission in grayscale overlaid
with contours. The contours are linear starting at  0.15 Jy beam$^{-1}$
and a step size of 0.10 Jy beam$^{-1}$. The left panel shows a cut along
the disk plane (p.a. = 58\degr{}), while the right panel shows a cut
along the outflow (p.a. = 145\degr{}). The systemic velocity is
indicated with a vertical line. The direction of the cuts is indicated
at top and bottom left of each figure. NGC\,7538\,S is at offset 0.}
\end{figure}

We have therefore made position velocity plots centered on the protostar
along the disk plane and in the outflow direction (Fig
\ref{fig-dcn-cuts}), i.e. the same cuts as we did for H$^{13}$CN. These
position velocity plots clearly demonstrate that DCN is strongly
affected by the outflow. The total velocity extent of the DCN emission
is $\sim$ 11 km~s$^{-1}$, which is far more than we would expect to see
from a rotating Keplerian disk. Furthermore, the cut along the outflow
shows that the blue-shifted emission is to the north-west and the red-shifted
emission is to the south-east of  NGC\,7538\,S.  The ``high-velocity''
DCN outflow is much more compact, only a few arcseconds , than the
outflow we see in HCO$^+$,  which extends $\sim$ 8\arcsec\ to the
northwest and even further in the red-shifted outflow lobe. However, at
low blue-shifted velocities, DCN has a similar extent as HCO$^+$. In the
cut along the disk, there is a clear blue-shifted velocity gradient
to the northeast, and red-shifted to the southwest. The red-shifted
emission to the southwest of the disk center is partly suppressed by 
red-shifted self-absorption, see Figure \ref{fig-spectra} and Table
\ref{tbl-3}, which quenches the emission on the red-side. There is
apparently a velocity component at $\sim$ $-$55 km~s$^{-1}$ extending
$\sim$10\arcsec\ to the southwest, which gives the appearance of a
velocity gradient across the disk. However, the velocity feature we
would associate with the disk, is the strongest red-shifted peak $\sim$
2\ptsec6 to the southwest at a velocity of  $-$53.5 km~s$^{-1}$.  The
corresponding blue-shifted peak is at 2\ptsec5 to the northeast  at a
velocity of $-$58.0 km~s$^{-1}$. There is, however another fainter even
more red-shifted (V$_{lsr}$ $\sim$ $-$51.5 km~s$^{-1}$) peak $\sim$ 2\arcsec\
to the southwest, which is probably from the outflow.

\begin{figure}[h]
\includegraphics[ scale=0.375,angle=-90]{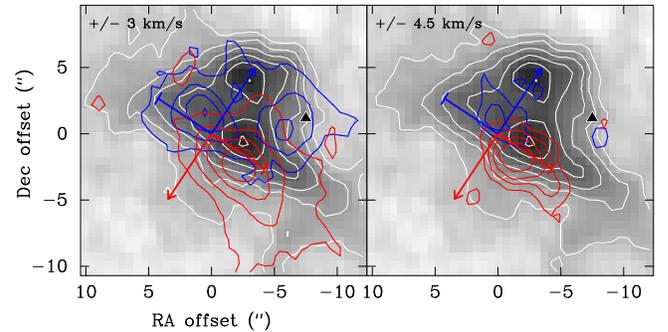}
\figcaption[]{
\label{fig-dcn-outflow}
Grayscale image of DCN \jtra32 emission overlaid with white contours
and integrated over 1.5 km~s$^{-1}$ (3 channels) centered on the cloud
velocity, $-$56~km~s$^{-1}$.  The greyscale image is overlaid with blue-
and red-shifted emission plotted with thick blue and red contours. In the
right panel we show emission integrated over 1.5  km~s$^{-1}$ and shifted
by 3 km~s$^{-1}$ relative to the cloud velocity, while the emission in
the right panel is shifted by 4.5 km~s$^{-1}$. The plane of the disk
is indicated by a line going through the center at a p.a. of 58\degr{}
and colored blue and red on the blue- and red-shifted side of the disk,
respectively. The position, extent, and direction  of the outflow is
indicated with arrows, blue to the northwest and red to the southeast.}
\end{figure}

To get a better idea of where the blue- and red-shifted emission comes
from, we overlaid the blue- and red-shifted DCN emission in two
different velocity intervals on the integrated DCN emission of the cloud
core (Figure \ref{fig-dcn-outflow}).  At a velocity offset of $\pm$3
km~s$^{-1}$  most of the blue-shifted emission is to the north
northwest, i.e. what one would expect if the emission comes from the
disk surface and/or the shearing layer between the outflow and the
surrounding cloud. There is also a clear north-east southwest
gradient, although with a somewhat different position angle, p.a. $\sim$
40\degr, than what we deduced from H$^{13}$CN.  At higher velocities,
$\pm$4.5 km~s$^{-1}$ the blue- and red-shifted DCN emission moves
more in the outflow direction, suggesting that most of the emission
originates from the outflow. In both velocity intervals, however, we see
very little red-shifted emission to the east. This could suggest that
some of the red-shifted emission is neither associated with the disk nor
the outflow from NGC\,7538\,S, but instead comes from more red-shifted
gas clumps in the cloud core.  

Since DCN \jtra32 is strongly affected by the outflow, it is not an
ideal disk tracer. If the dominant velocity peaks seen in the position
velocity plot along the disk plane at $\sim$ $\pm$2\ptsec5 originate
from a rotating disk, the DCN emission would suggest a systemic velocity
of $-$55.8 km~s$^{-1}$, a velocity shift of 4.5 km~s$^{-1}$ for a disk
size of  5\ptsec1, i.e. very similar in size to what we get from
H$^{13}$CN, but with a larger velocity difference.  We can also see this
in Figure \ref{fig-disk-spectra}, where we show spectra of both
H$^{13}$CN and DCN in three positions of what we interpret as the
rotating accretion disk surrounding NGC\,7538\,S. Both molecules are
blue-shifted to the northeast, both show some evidence for slightly
red-shifted narrow self-absorption towards the center of the disk and
red-shifted emission to the southwest,. Here we have ignored the strong,
relatively narrow emission at $-$56 km~s$^{-1}$, which is seen in DCN,
but which is far less evident in H$^{13}$CN. The velocity gradient is
larger in DCN than in H$^{13}$CN, which could result from the higher
angular resolution  in DCN compared to H$^{13}$CN.

\subsubsection{ CH$_3$CN J = 12 $\to$ 11}
\label{mcn}

\begin{figure}[b]
\includegraphics[ scale=0.6]{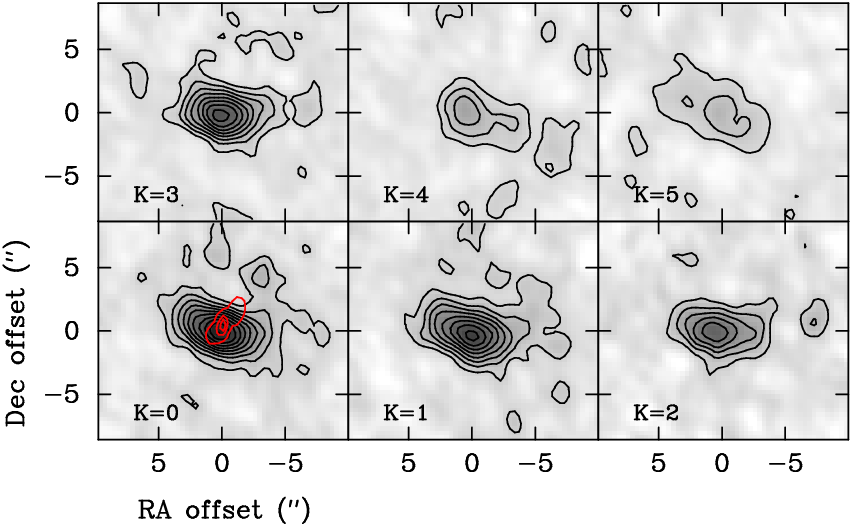}
\figcaption[]{
\label{fig-mcn-disk}
BIMA contour images of integrated CH$_3$CN J = $12 \to 11$ for K-levels
0 to 5. For each K-level we have integrated the line emission over a 4.2
 km~s$^{-1}$ wide velocity interval, except for K=4 and 5, where we have
used a wider velocity interval, 5.3  km~s$^{-1}$. The offsets are
relative to \mbox{$\alpha_{2000.0}$ = 23$^h$ 13$^m$ 44\psec98},
\mbox{$\delta_{2000.0}$ = $+$61\degr{} 26\arcmin{} 49\ptsec5}, which is
0\farcs3 to the south of the protostar. The beam FWHM is 2\ptsec76
$\times$ 2\ptsec36  PA 22.9 \degr.}
\end{figure}

\begin{deluxetable}{lccr}
\tablecolumns{4}
\tablenum{5}
\tablewidth{0pt} 
\tablecaption{Line parameters for CH$_3$CN  \jtra{12}{11}  towards the center of  the accretion disk}
\label{tbl-5}
\tablehead{
\colhead{K level} & \colhead{$V_{\rm LSR}$} &  \colhead{$\Delta V$} & \colhead{$\int T_{\rm MB} dV$}\\
\colhead{  }  &\colhead{ [km s$^{-1}$] }&\colhead{ [km s$^{-1}$] } &\colhead{ [K~km s$^{-1}$] }\\
}

\startdata
\sidehead{HPBW = 2.76 $\times$ 2.36\tablenotemark{a}}
\cline{1-4}
0        &   $-$56.4 $\pm$ 0.1 \tablenotemark{b} &  5.35  $\pm$ 0.17 &  16.6  $\pm$ 0.9\\
1        &   \nodata &  \nodata & 11.6 $\pm$  0.8\\ 
2        &  \nodata &  \nodata  & 9.4 $\pm$ 0.8 \\ 
3        & \nodata & \nodata & 12.0 $\pm$ 0.9\\
\sidehead{HPBW = 3.16  $\times$ 2.42\tablenotemark{a}}
\cline{1-4}
4     &  $-$56.9  $\pm$ 0.5  &  6.2   $\pm$ 1.3   & 5.4 $\pm$ 0.9\\
5     &  $-$55.3 $\pm$  1.3  &  7.3  $\pm$ 4.1    & 4.3 $\pm$ 1.6 \\
6     &  $-$56.6 $\pm$  0.4  &  1.8  $\pm$ 1.2    & 1.1 $\pm$ 0.4 \\
\sidehead{CH$_3^{13}$CN}
\cline{1-4}
2 &  $-$56.3  $\pm$ 1.6  &  7.5  $\pm$ 3.5  & 2.7 $\pm$ 1.1\\
3 &  $-$55.8  $\pm$ 0.6  &   2.8  $\pm$ 1.2  & 1.2 $\pm$ 0.5\\
\cline{1-4}
\enddata

\tablenotetext{a}{Image containing K-levels 0 to 3 have better uv-coverage
than images containing K levels 4, 5, and  6, and the K = 2 and 3
transitions of CH$_3^{13}$CN. The latter data set has instrumental problems
causing the data to be excessively noisy.} \tablenotetext{b}{CLASS fit
by keeping velocity separations between K-levels to their theoretical
values and forcing the line width to be the same for all transitions.}
\end{deluxetable}

In this section we discuss the structure, kinematics, opacity and
temperature of CH3CN.  We show that although there is CH$_3$CN emission
from the accretion disk, emission from the outflow dominates and hence
prevents CH$_3$CN from being a good disk tracer for NGC\,7538\,S.

The emission from CH$_3$CN J = 12 $\to$ 11 is confined to the immediate
vicinity of NGC\,7538\,S (Figure \ref{fig-mcn-disk}). The emission is
well resolved and shows an elliptical source centered at 0\psec2 $\pm$
0\psec03, $-$0\ptsec4 $\pm$ 0\ptsec1, relative to our nominal position
for NGC\,7538\,S, i.e. slightly south of the stellar position. At  K =
0, which has an excitation energy of  68.9 K above ground,  the size of
the CH$_3$CN emission is $\sim$ 4\ptsec9 $\times$ 1\ptsec5 with a p.a.
of 70\degr. The emitting region gets progressively smaller for higher K
levels, suggesting that the gas is hotter closer to the star. At K = 3,
which has an excitation energy of 132.9 K, the size of the emitting
region 4\ptsec0 $\times$ 1\ptsec5 with a p.a. of 84\degr. At higher
K-levels the emission is faint and the source size more uncertain, but
appears to follow the same trend seen for lower K-levels, i.e. the
emission becomes more compact and we see a gradual increase in the
position angle. It would therefore appear that  CH$_3$CN traces the
accretion disk, although at low K-levels the position angle of the
emission differs from what what we observe in dust continuum and
H$^{13}$CN. At higher K-levels the CH$_3$CN emission becomes even more
misaligned, which we would not expect to observe if the emission
originates in the accretion disk, unless the inner disk is warped
relative to the colder, more extended disk. As we will see
below, the CH$_3$CN emission is dominated by the outflow.

\begin{figure}[t]
\includegraphics[ scale=0.75,angle=-90]{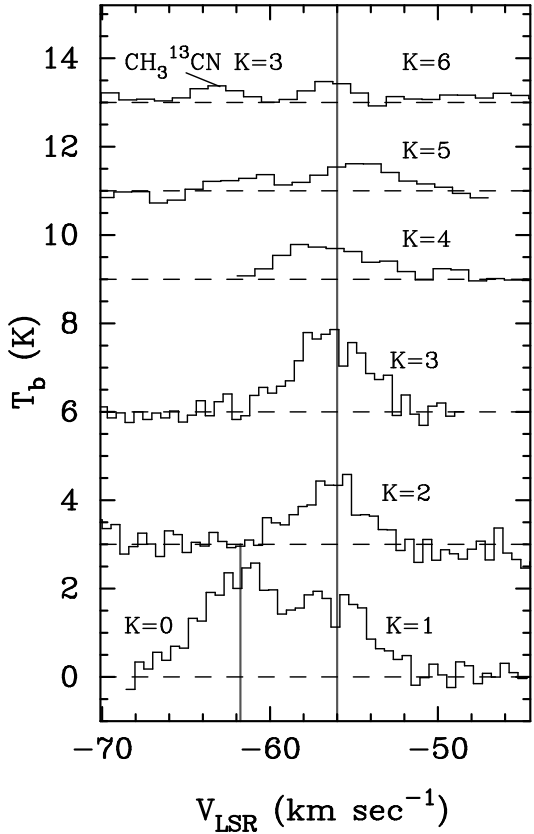}
\figcaption[]{
\label{fig-mcnspec}
Observed CH$_3$CN J = $12 \to 11$ spectra towards the center of
NGC\,7538\,S for K = 0 to 6. The grey vertical line indicates the
systemic velocity of the protostellar disk. The K = 0 and K = 1 spectra
are blended, which is seen in the bottom spectra. The velocity corresponds
to the K = 1 level, but we have also indicated with a grey line where the
systemic velocity falls for the K = 0 line.  The top spectrum includes
the CH$_3$$^{13}$CN K = 3 line, which  is separately labeled in the
panel. The peak ratio of CH$_3$CN to  CH$_3$$^{13}$CN for K = 3 is $\sim$
3 - 10, indicating that the line is optically thick. Self absorption at the
systemic velocity of the cloud or at slightly red-shifted velocities is
seen in K-levels 0 to 3.}
\end{figure}

\begin{figure}[t]
\includegraphics[ scale=0.375,angle=-90]{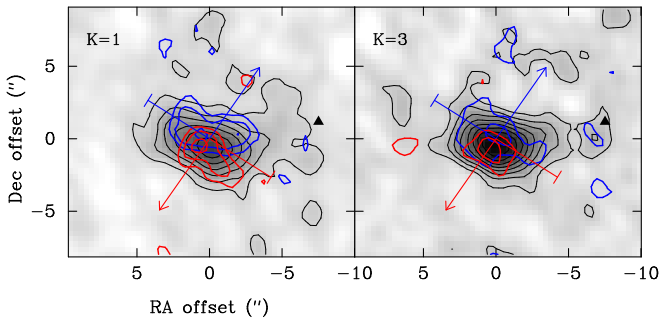}
\figcaption[]{
\label{fig-k012do}
Greyscale plots enhanced with  contours of integrated  CH$_3$CN J
= $12 \to 11$ K = 1  emission (left panel) and K=3 emission (right
panel). Both K=1 and K=3 are integrated over 4.2 m~s$^{-1}$ as in Fig
~\ref{fig-mcn-disk}. These contour plots are overlaid with contours
of integrated high velocity emission, plotted in blue and red for the
blue-shifted and red-shifted wings of the emission. The integrated
blue- and red-shifted emission is aligned along the outflow direction
(p.a. $\sim$ 145\degr, demonstrating that CH$_3$CN is tracing the outflow,
not the disk. }
\end{figure}

\begin{figure}[t]
\includegraphics[ scale=0.34,angle=-90]{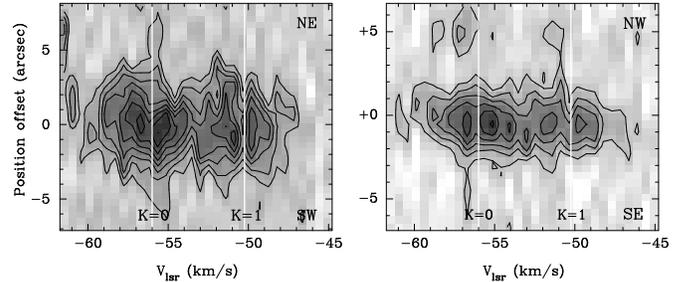}
\figcaption[]{
\label{fig-k01-cuts}
Position velocity cuts of  CH$_3$CN J = $12 \to 11$ K = 0 and K=1 emission
along the disk plane (left panel) and along the outflow (right panel).}
\end{figure}

In Figure \ref{fig-k012do}  we plot the integrated red- and blue-shifted
emission of the K = 1 and 3 transitions and overlay them on the
integrated emission from the line core of these lines, shows that the
CH$_3$CN emission is dominated by the outflow. This is even more
striking for the K = 3 transition, where the blue- and red-shifted line
emission is well separated and aligned with the outflow. The cut along
the disk-plane (Figure \ref{fig-k01-cuts}), however, shows only a
marginal velocity gradient over the disk, yet, as we already seen, both
H$^{13}$CN \jtra{1}{0} and DCN \jtra{3}{2} show a strong velocity
gradient over the disk. Although this is somewhat surprising, it does
not contradict our earlier results. CH$_3$CN  is strongly dominated by
the outflow, which may mask the emission from the accretion disk.

It is also possible that the CH$_3$CN emission is optically thick, which
will make it difficult to see rotation of the disk.  CH$_3$CN  spectra
towards the center of NGC\,7538\,S are shown in Figure
\ref{fig-mcnspec}.  We detected all the K transitions, K = 0 to 6, and
the K = 2 and 3 transitions of  CH$_3$$^{13}$CN, which were included in
our frequency setting (Table \ref{tbl-2}).  In Table \ref{tbl-5} we give
results from Gaussian fits to the spectra presented in Figure
\ref{fig-mcnspec}. We fitted the K = 0 to 3 levels simultaneously by
keeping the separation between the K components to their theoretical
value, but allowed the FWHM to be a free parameter and constrained to be
the same for all these K values.  The K-levels 4 to 6 were fitted
individually. It is clear from these fits that the line widths gets
broader for increasing K-value, which is what one would expect, since
these lines probe the hot outflow. The line profiles are not well fitted
with a single Gaussian, because they have a broader underlying pedestal
from the outflow.  If we fit the K = 3 transition with a two-component 
Gaussian, we find that the integrated intensity of the broader
component, $\Delta$v = 9.3 km~s$^{-1}$, is almost twice the value of the
narrow (disk-)component. There is also a hint of a narrow
self-absorption feature in the K = 0 to K = 3 transitions
(Figure~\ref{fig-mcnspec}), suggesting that there is cold, somewhat
optically thick CH$_3$CN gas in the surrounding cloud core. This
self-absorption is also seen in the position velocity plots
(Figure~\ref{fig-k01-cuts}), confirming that it is real and not an
artifact. Narrow self-absorption is also seen in the K = 3 transition,
but slightly more red-shifted. 

Since we detected emission from the K = 2 and 3 transitions of
CH$_3$$^{13}$CN, we can use the line ratios to check whether  CH$_3$CN
is optically thin or thick.   The ratios of the integrated
CH$_3$CN/CH$_3$$^{13}$CN line intensities are about 3 - 10, which
strongly suggests that the CH$_3$CN lines are optically thick. Although
CH$_3$CN is often optically thin in hot, dense molecular cloud cores,
see e.g. \citet{Araya05}, it is not uncommon to find optically thick 
CH$_3$CN, especially in high-mass star forming regions
\citep{Pankonin01,Watson02,Remijan04,Beltran05}, when they are observed
with high spatial resolution.

\begin{figure}[t]
\includegraphics[ scale=0.47,angle=-90]{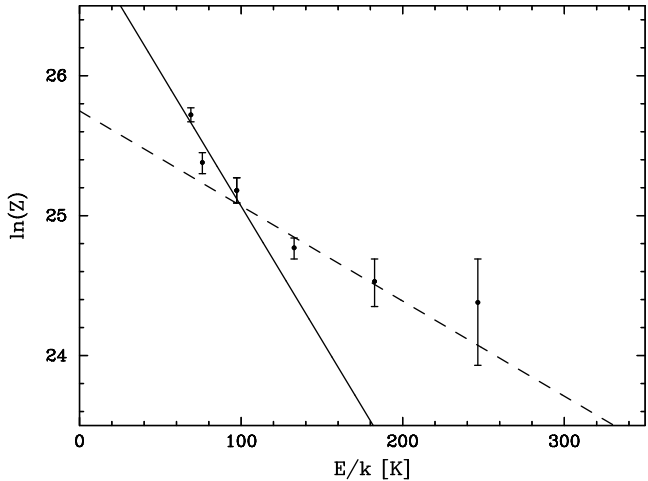}
\figcaption[]{
\label{fig-mcn_rot}
Population-diagram of  CH$_3$CN \jtra{12}{11} towards NGC\,7538\,S, see
discussion in Section \ref{mcn}. The solid curve shows a least squares
fit to the three lowest K-transition, K = 0 to K = 2, while the dashed
line is a fit for K = 2 to 5. The low K-transitions, K = 0 to 2, for
which the emission is likely to include a non-negligible contribution
from the accretion disk, correspond to a temperature of $\sim$ 50 K,
while the high K-transitions, which are dominated by the outflow,
correspond to a temperature of 150 K. }
\end{figure}

CH$_3$CN is traditionally analyzed using a rotational temperature
equilibrium (RTE) analysis \citep[and references
therein]{Loren84,Araya05}. However, we know that some of the 
K-transitions are optically thick and that the source size changes as a
function of K-level. Furthermore, we also know that we have some
contribution from both the disk and the outflow, especially at low
K-levels. An RTE or  population-diagram analysis assumes that the gas is
optically thin and  in LTE at a single temperature, none of which is
strictly true in this case. Nevertheless, it is instructive to
look at the results of such an analysis. Since \citet{Araya05} have done
a recent in  depth discussion of population-diagram analysis for
CH$_3$CN, there is no reason for us to repeat any of the discussion
here. Following \citet{Araya05} we do a least squares fit linear fit of
the equation

\begin{displaymath}
ln (Z) = ln \biggr(\frac{N_{CH_3CN}}{Q(T_{rot})}\biggl)  -  \frac{E_{JK}}{kT_{rot}}
\end{displaymath}

\noindent where  Z = $\frac{N_{JK}}{g_{JK}}$. To minimize the effect
from the change in source size, we use the measured antenna temperatures
for the peak position of  CH$_3$CN ignoring any difference due to
coupling of the source, except for a small correction for the K = 4 and
K = 5 levels , which were imaged with a slightly different beam size
(Table \ref{tbl-5}). Here  we took the square of the ratio of the beam
size, resulting in a correction factor of 0.83. The results of this
analysis is shown in Figure \ref{fig-mcn_rot}. The population-diagram
indicates that the CH$_3$CN emission cannot be characterized with a
single rotational temperature. We have therefore separately done a least
squares fit to the three lowest K-transitions, K = 0 to 2, for which the
emission is likely to include a non-negligible contribution from the
accretion disk, and for K = 2 to 5, for which we seen that the outflow
is dominates. The fit to the low K-transitions give T$_{rot}$ = 52 $\pm$
10 K, while the high K-transitions give T$_{rot}$ = 147 $\pm$ 40 K.
Neither of these temperatures are very accurate. The low K-transitions
are definitely optically thick, which will lead to an overestimate of
the rotational temperature. Contribution from the outflow will also
raise the temperature. The temperature of the disk is certainly less
than 50 K and higher than 25 K, the temperature of the surrounding cloud
core, since we see some self-absorption from virtually every molecule
tracing the disk. The temperature of the outflow is not well constrained
 by our observations, because we have poor signal-to-noise at high
K-levels, which could more accurately constrain the temperature of the
outflow. However, analysis of the two H$_2$CO transitions 3$_{03}  \to
2_{02}$ and 3$_{22}  \to 2_{21}$, which are dominated by the outflow,
also indicate a temperature of $\geq$ 100 K, in good agreement with what
we derive from CH$_3$CN. \citet{Kalenskii00}, who did single dish
observations of  NGC\,7538\,S in the \jtra65 and \jtra54 transitions of
CH$_3$CN with a beam size of $\sim$40\arcsec,  derived a rotation
temperature of $\sim$ 40 K, much lower than what we observe. They also
measure a different radial velocity, -55.4 km~s$^{-1}$, while we find a
radial velocity of $\sim$ -56.3  km~s$^{-1}$ (Table~\ref{tbl-5}). Since
their beam covers most of the cold cloud core surrounding NGC\,7538\,S,
it appears that the CH$_3$CN emission they observe is dominated by the
cloud core, which we filter out in our BIMA observations.

\section{Discussion}

\subsection{Is NGC\,7538\,S surrounded by a rotating accretion disk?}
\label{rotating_disk}

NGC\,7538\,S powers a highly collimated thermal jet and drives a very
young, hot molecular outflow, suggesting that it must be surrounded by a
rotating accretion disk. We have shown that
the star is embedded in a compact elliptical cloud core, which is
approximately perpendicular to the thermal jet and the associated
molecular outflow. High angular resolution observations  tracers like 
H$^{13}$CN \jtra10, HN$^{13}$C \jtra10, H$^{13}$CO$^+$ \jtra10,
and DCN \jtra32, which all are largely optically thin,  show 
blue-shifted line emission to the northeast of NGC\,7538\,S and red-shifted to the
southwest, with the highest velocities towards the center of the
protostar. Such a velocity  signature is characteristic for a
rotating Keplerian disk. However, some of the velocity broadening seen
towards the center of the protostar might be caused by the outflow
powered by the disk, instead of originating in the faster rotating inner
part of the disk. We therefore have to review our observational results
with caution. We know that the outflow is much hotter than the bulk of
the gas in the disk. It therefore tends to dominate the emission even
for molecules, which are normally believed to be good disk tracers like
CH$_3$CN \citep{Cesaroni97,Cesaroni05,Beltran04,Beltran05,Patel05,Furuya08}.
In Section~\ref{mcn} we show that for NGC\,7538\,S the CH$_3$CN emission
originates from the dense outflow gas near the surface of the disk,
where the outflow is launched. The emission from CH$_3$CN therefore
appears roughly orthogonal to the outflow.

Based on our discovery of a Keplerian like rotation in NGC\,7538\,S
using H$^{13}$CN \jtra10 \citep{Sandell03},  we thought that DCN might
work equally well and would provide us higher spatial resolution, if we
observed the \jtra32 transition at 1.4 mm. However, as we saw in Section
\ref{dcn}, the DCN \jtra32 transition is clearly affected by the
outflow. Even though it also traces the dense gas in the disk, the
relatively strong emission from the outflow makes  it  difficult to
reliably separate the outflow emission from that of the disk. For
estimating a Keplerian mass, we therefore only use the results from
H$^{13}$CN \jtra10, because  H$^{13}$CN  appears least affected by the
outflow, and therefore provides the best information about the
kinematics of the neutral gas surrounding NGC\,7538\,S. However, because
NGC\,7538\,S is surrounded by a very dense cold cloud core,  all our
high density tracers, including H$^{13}$CN, are affected by
self-absorption. Most of this self-absorption is likely to originate in
cold dense gas in the surrounding  cloud envelope, but some of it could
also originate in the colder outer portions of the disk, because the
self-absorption becomes much more prominent at the center of the disk at
the highest spatial resolutions.  Since the self-absorption is slightly
red-shifted (Figure~\ref{fig-h13cncuts}, right panel), and almost
certainly dominated by the accretion flow towards the central protostar,
it makes the line emission appear more blue-shifted. Southwest of
NGC\,7538\,S there is also an extended cloud component roughly at the
systemic velocity of the cloud, see Section \ref{dcn} and Figure
\ref{fig-dcn-chann}. This emission blends in with the red-shifted
emission from the rotating disk, which together with an unknown amount
of red-shifted self-absorption results in uncertain velocities in the
red-shifted side of the disk.

Despite these difficulties, our new observations clearly confirm a
velocity gradient in the dense gas disk surrounding NGC\,7538\,S.
Position velocity plots of H$^{13}$CN \jtra10 and DCN \jtra32 show a
roughly Keplerian like rotation curve centered on the protostar with a
diameter of $\lesssim$ 5\arcsec{} (14,000 AU). If we assume that the
disk is rotationally supported and take the velocities derived from
H$^{13}$CN,  we derive an enclosed mass of  $\sim$ 14 \Msun, uncorrected
for inclination. The inclination angle of the disk is $\sim$ 50\degr.
Therefore the enclosed mass of the disk and the central protostar is
$\sim$ 24 \Msun. We do not have a mass estimate for the central
protostar, but based on the bolometric luminosity, $\sim$ 10$^4$ \Lsun,
it is probably has a mass of $\sim$ 10\Msun, or about the same order as
the mass of the disk.  Such massive disks can be self-regulated at a
condition of  marginal Jeans stability and still appear to have an
approximately Keplerian rotation curve \citep{Bertin99}.

\citet{Cesaroni07} argue that disks like the one \citet{Sandell03}
discovered around NGC\,7538\,S should be called toroids rather than
disks. They further argue that such toroids may simply be transient
structures, which may or may not evolve into a ``classical'' Keplerian
disk. In this follow-up study we show that the Keplerian disk
surrounding NGC\,7538\,S, is much smaller than previously  thought and
it is therefore not a  transient structure. In high-mass protostars, the
extreme accretion rates of $\geq$ 10$^{-4}$ M$_{\odot}$~yr$^{-1}$,  and
the high pressure from the surrounding cloud, may enable  such a 
``disk'' to survive long enough to form an O-star. In this respect,
NGC\,7538\,S, could well be a younger analogue of NGC\,7538\,IRS\,1,
which by all accounts still appears to be surrounded by an accretion
disk \citep{Sandell09}. In the case of IRS\,1, which has a luminosity of
an O7 star, but which is still heavily accreting, the disk has probably
evolved into a more ``classical'' thin disk, which is rather hard to
image, because the free-free emission from the central O-star is much
brighter than the disk at mm-wavelengths, which is the only wavelength
regime, where we have arrays with sufficient spatial resolution to
potentially resolve and image such a disk.

\subsection{Is NGC\,7538\,S a high-mass star?}

Our observations confirm that NGC\,7538\,S lies in the center of a cold,
massive cloud core with a diameter $\sim$ 1 pc, and a mass 2,000 \Msun
(Section \ref{core}), which therefore provides the necessary conditions
for the formation of high-mass stars, i.e. essentially the whole core
has a surface density, $\Sigma$ $\gtrsim$  1 g~cm$^{-2}$
\citep{McKee02,Krumholz08}.  NGC\,7538\,S coincides within errors with a
cold far-infrared source with a luminosity $\sim$ 1.5 10$^4$ \Lsun\
\citep{Werner79,Thronson79}, suggesting an early B-star, if it is a
single star. IRS\,11, the only near-IR source in the vicinity of
NGC\,7538\,S, can only account for about a tenth of this luminosity, and
the same is true for the small cluster of Spitzer 8 $\mu$m sources
surrounding IRS\,11 (Paper II). Therefore the bulk of the luminosity
appears to be generated in the very massive elliptical cloud core
($\sim$ 100 \Msun{}) cloud core, in which NGC\,7538\,S is embedded. Here
NGC\,7538\,S completely dominates the luminosity. It drives a highly
collimated thermal jet (Paper II), powers a massive, very young bipolar
outflow \citep{Sandell03}, and excites OH, CH$_3$OH (Class II), and
H$_2$O masers \citep{Argon00,Pestalozzi06,Kameya90}. OH 1665-MHz masers
are only found in young massive star forming regions
\citep{Caswell98,Argon00}, mostly associated with Ultra Compact
\ion{H}{2} regions.  The same is true for Class II CH$_3$OH masers
\citep{Minier03}, i.e. no CH$_3$OH maser have yet been found toward a
low or intermediate mass star. Although CH$_3$OH masers sometimes are
associated with OH masers, the majority of them are found in weak or
radio quiet regions. They are therefore believed to trace a younger,
perhaps protostellar phase, in the evolution of high-mass stars
\citep{Caswell98,Ellingsen06}. The observed accretion rate for the disk
surrounding NGC\,7538\,S is $\sim$  1 10$^{-3}$ M$_{\odot}$~yr$^{-1}$.
Such accretion rates have not been seen in low or intermediate mass
protostars, but  accretion rates in the range 10$^{-3}$ - 10$^{-4}$
M$_{\odot}$~yr$^{-1}$ are commonly inferred for young high-mass stars of
comparable luminosity
\citep{Molinari98,Cesaroni99,Beuther02b,Beuther03,Fontani04}. Accretion
rates as high as a few times 10$^{-2}$ M$_{\odot}$~yr$^{-1}$ have been
reported \citep{Beltran04,Beltran05}, although such high accretion rates
are generally only seen towards high luminosity UC \ion{H}{2} regions
\citep{Hofner03,Zapata08}. Theoretical models of collapsing clouds show
that the accretion rates increase rapidly in the beginning of the
collapse phase and decrease monotonically later on \citep{Li98,Yorke02},
which may be why the observed accretion rate is on the high side for a
region which has a total luminosity of $\sim$ 1.5 10$^4$ \Lsun. Since
NGC\,7538\,S has the luminosity of a high-mass star and excites OH and
CH$_3$OH Class II maser emission, which has only been seen in high-mass
stars, it must be a high-mass star.

\section{Summary and Conclusions}

We have carried out extensive observations of the star forming core
NGC\,7538\,S with BIMA in mostly optically thin tracers with  spatial
resolutions ranging from $\sim$ 3\arcsec\ to 8\arcsec. Additionally we
have acquired complementary observations with FCRAO and JCMT to fill in
lacking short spacing in our BIMA observations, which are needed to
improve the image fidelity and  reliability of our images of molecules
like HCO$^+$, H$_2$CO, and H$^{13}$CN, which are spatially very
extended.

We confirm that there is a very young high-mass (proto)star in the
compact elliptical core, which has a size of 8\arcsec\ $\times$
3\arcsec\ and a mass of  85 - 115 \Msun. Recent sub-arcsecond continuum
imaging at 110 and 224 GHz with CARMA (Paper II) resolve the elliptical
core into three compact sources, all of which are almost certainly
protostars.  The strongest one of the three sources agrees within
0\farcs15 with the adopted position for the high-mass protostar, which
is $\sim$ 2\arcsec\ to the northeast from the center of the compact
elliptical core. The protostar is seen as a faint, extremely obscured
mid-IR source, which coincides with a VLA thermal jet, and an OH and a
CH$_3$OH class II methanol maser. The star is surrounded by a massive,
rotating accretion disk, which drives a highly collimated thermal jet
and powers a very compact,  hot molecular outflow. We see clear evidence
for accretion towards the protostar. Almost all molecular transitions
that we have observed, show red-shifted self-absorption, which can be
explained only by infall motions of the gas in front of the protostar.
The accretion signature is very strong in the optically thick HCO$^+$
\jtra10 transition, which shows infall velocities up to $\sim$ 15
km~s$^{-1}$. We have used this accretion profile to derive a direct
estimate of the accretion rate towards the disk, and find an accretion
rate $\sim$ 10$^{-3}$ M$_{\odot}$~yr$^{-1}$. We also estimated the
accretion rate by assuming that the accretion rate is about three times
the outflow rate, and by assuming a steady state infall to the disk,
with an  infall velocity equal to the observed rotation velocity of the
disk at the radius R.  All three methods give very similar results. Such
high accretion rates are sufficient to quench the formation of an
\ion{H}{2} region and allow the central protostar to continue to grow in
mass \citep{Walmsley95,Keto03,Keto07}.

The rotating accretion disk is best seen in H$^{13}$CN \jtra10.
H$^{13}$CN is only marginally affected by emission from the intense hot
outflow, while the emission from the outflow may dominate or severely
hinder us from seeing the disk in many of the molecular transitions that
we have observed. We found that DCN  \jtra32, which we expected to be
an equally good disk traces as  H$^{13}$CN, is strongly affected by the outflow. 
Due to the high angular resolution, $\sim$ 2\ptsec6 (Table \ref{tbl-2}), 
we can separate most of the outflow
emission from the disk-emission and therefore confirm that NGC\,7538\,S
is surrounded by a rotating accretion disk. The emission from
methylcyanide, CH$_3$CN, however, is largely optically thick and the
higher K-levels  are completely dominated by the hot outflow emission near
the surface of the disk.

BIMA observations of  H$^{13}$CN \jtra10 supplemented with FCRAO data
show that the cloud core, in which NGC\,7538\,S is embedded, has a radius
of $\sim$ 0.5 pc, and a mass $\sim$ 2000 \Msun. The size and mass of
the core agrees very well with what we derive from analysis of the
SCUBA 850 and 450 $\mu$m data presented by \citet{Sandell04}, while the
mass estimate from a single dish  C$^{18}$O \jtra21 map is much lower,
suggesting that CO is depleted (frozen onto grains) in the cold cloud
envelope. The cloud core is  dense and massive enough to provide the
necessary conditions for high-mass star formation.

\acknowledgements The
BIMA array was operated by the Universities of California (Berkeley),
Illinois, and Maryland with support from the National Science
Foundation. We want to thank  Dr. W. M. Goss for helpful comments and 
support throughout this project. Special thanks goes to Dr. Mark Heyer, 
who did the FCRAO observations for us, and to the JCMT telescope system 
specialists, who did all the JCMT observing in service mode.  We  thank the 
anonymous referee for an extremely careful reading of our manuscript, which 
considerably improved our paper.

{}

\end{document}